\documentclass[10pt]{iopart}
\usepackage{graphicx}
\usepackage{epsfig}
\usepackage{iopams}
\usepackage[usenames]{color}
\usepackage{ifthen}
\usepackage[T1]{fontenc}
\usepackage[utf8]{inputenc}
\usepackage{times}
\usepackage{color}
\usepackage{cite}
\usepackage{cuted}

\def\be{\begin{equation}}
\def\ee{\end{equation}}
\def\bea{\begin{eqnarray}}
\def\eea{\end{eqnarray}}
\def\bi{\begin{itemize}}
\def\ei{\end{itemize}}

\newcommand{\op}[1]{\widehat{#1}}
\newcommand{\dagop}[1]{\widehat{#1}^{\dagger}}
\newcommand{\bo}[1]{{\mathbf{#1}}}
\newcommand{\mc}[1]{{\mathcal{#1}}}
\newcommand{\wt}[1]{{\widetilde{#1}}}
\newcommand{\wb}[1]{{\overline{#1}}}
\newcommand{\nonu}{\nonumber}

\newcommand{\bra}[1]{\left\langle#1\right\vert}
\newcommand{\ket}[1]{\left\vert#1\right\rangle}
\newcommand{\braket}[2]{\langle#1\vert#2\rangle}

\newlength{\templength}

\newcommand{\half}{\ensuremath{\frac{\scriptstyle 1}{\scriptstyle 2}}}

\newcommand{\text}[1]{{\rm{#1}}}

\newcommand{\REM}[1]{\ifthenelse{1==0}{#1}{}} %not printed

\begin{document}

\title{Quantum fluctuation effects on the quench dynamics of thermal quasicondensates}

\author{Tomasz \'Swis\l{}ocki and Piotr Deuar}
\address{Instytut Fizyki PAN, Aleja Lotnik\'ow 32/46, 02-668 Warsaw, Poland}
\ead{deuar@ifpan.edu.pl}

%\author{Piotr Deuar}
%\affiliation{Instytut Fizyki PAN, Aleja Lotnik\'ow 32/46, 02-668 Warsaw, Poland}

\begin{abstract}
We study the influence of quantum fluctuations on the phase, density, and pair correlations in a trapped quasicondensate after a quench of the interaction strength.
To do so, we derive a description similar to the stochastic Gross-Pitaevskii equation (SGPE) but keeping a fully quantum description of the low-energy fields using the positive-P representation. 
This allows us to treat both the quantum and thermal fluctuations together in an integrated way. A plain SGPE only allows for thermal fluctuations. 
The approach is applicable to such situations as finite temperature quantum quenches, but not equilibrium calculations due to the time limitations inherent in positive-P descriptions of interacting gases.
One sees the appearance antibunching, the generation of counter-propagating atom pairs, and increased phase fluctuations. We show that the behavior can be estimated by adding the $T=0$ quantum fluctuation contribution to the thermal fluctuations described by the plain SGPE.
\end{abstract}

%\submitto{\jpb}
\maketitle

\ioptwocol

%%%%%%%%%%%%%%%%%%%%%%%%%%%%%%%%%%%%%%%%%%%%%%%%%%%%%%%%%%%%%%%%%%%%%%%
\section{Introduction}
%%%%%%%%%%%%%%%%%%%%%%%%%%%%%%%%%%%%%%%%%%%%%%%%%%%%%%%%%%%%%%%%%%%%%%%

Fluctuations of observed quantities in many-body quantum systems arise in a variety of ways. Two classes of a distinctly different nature are: {\it thermal} fluctuations due to successive observations being made on different 
components of the mixture that is the thermal ensemble, and the so-called {\it quantum} fluctuations that arise as a consequence of the observation itself. An interacting many-body state is rarely, if ever, in an eigenstate of few-body observables such as densities or correlations, so that a randomness appears when these are measured. Such quantum fluctuations are present already in the $T=0$ ground state. In  ultracold gases they are related to  effects such as quantum shot noise, the quantum depletion of a condensate, production of atom pairs, or spontaneous scattering into empty modes. At nonzero temperatures, the two kinds of fluctuations coexist, and both contribute to observations. 

To include quantum fluctuations other than possibly simple shot noise, one must move beyond the mean field description of the Gross-Pitaevskii equation (GPE) that treats each atom as occupying the same orbital. 
At very low temperatures, they can be described well by Bogoliubov theory. This separates the system into one condensate mode that accounts for the vast majority of atoms and the remaining excited modes which are treated in a fully quantum manner but do not interact \cite{Bogoliubov47,Gardiner97,Castin98}. Some extensions have included back-action onto the condensate \cite{Castin98,Sinatra00,Gardiner07,Billam13}. This approach treats both quantum and thermal fluctuations on an equal footing. Unfortunately, the assumptions break down when the condensate fraction $n_0$ deviates appreciably from 100\% (as a rule of thumb, when $n_0\lesssim0.9$). 
At higher temperatures, the c-field methods, that treat the system as being composed of a number of relatively low-energy modes described individually by classical complex fields \cite{Davis01,Goral01,Bradley05,Connaughton05,Stoof99,Berloff02}, have been very successful (and reviewed in \cite{Brewczyk07,Blakie08,Proukakis08}). 
However, c-fields completely discard the quantum fluctuations in the treated modes, which makes them incapable of properly describing such effects as spontaneous scattering, pair formation, or quantum depletion, even at the low temperatures that are appropriate for Bogoliubov theory. 

An important question, then, is how and under what conditions do quantum fluctuations appreciably change the picture obtained with c-fields? Here we wish to make new inroads into these matters. 
What will be done is to take the master equation for the low-energy degenerate boson field that has been used to obtain the c-field SGPE description \cite{Gardiner03}, but then describe it in a fully quantum manner with the positive-P representation (PPR), rather than making the classical approximation. 

The Stochastic Gross-Pitaevskii equation (SGPE) \cite{Stoof99,Stoof01,Gardiner03,Bradley05,Proukakis08,Cockburn09} is a c-field description of the dynamics that has been used for a wide range of problems where thermal fluctuations are important. These include condensate growth \cite{Stoof01,Duine01}, defect formation \cite{Liu14}, soliton dynamics \cite{Cockburn11b}, and phase fluctuations \cite{Cockburn09,Cockburn11a,Wright10,Gallucci12}. While the quantum fluctuations in the c-field modes are disregarded, an approximate description of the low-occupation modes is incorporated in the form of a thermal bath, which is not a feature of most other c-field approaches. A convenient feature of the SGPE is that the temperature of the system can be imposed directly on the equations rather than determined post-fact on the basis of the properties of the Bose field \cite{Brewczyk07}.

The positive-P representation (PPR) \cite{Drummond80,Deuar06} is a full mapping of the quantum state and dynamics of the system onto a distribution of phase-space variables that then evolve stochastically. It has been used for simulating e.g. pair scattering and nonclassicality during condensate collisions \cite{Deuar07,Perrin08,Ogren09,Jaskula10,Kheruntsyan12,Deuar13,Lewis-Swan14,Deuar14,Lewis-Swan15}, condensate growth \cite{Drummond99} or fiber soliton dynamics\cite{Carter87,Drummond93,Corney06}, where the essence of the problem lies in correctly treating spontaneous scattering into a great number of empty modes. Its advantage over more direct fully quantum methods is that the numerical effort scales well (even linearly) with the size of the numerical lattice. It also readily allows for arbitrary trapping potentials or a time dependence of the Hamiltonian parameters. The reason that one cannot use the PPR directly in general cases is because of a nonlinear amplification of the noise in the equations that limits the time over which dynamics can be simulated \cite{Deuar06}. In particular, it is not generally possible to simulate long enough to reach the equilibrium state. 

A number of past works have, under various conditions, incorporated spontaneous processes in thermal gases that were not amenable to the standard Bogoliubov treatment. A notable one is the quasicondensate extension of Bogoliubov theory by Mora and Castin \cite{Mora03} which relies on small density fluctuations. The truncated Wigner method \cite{Steel98,Sinatra00,Sinatra02,Norrie05b,Polkovnikov10} has been widely used, one example being the thermal decay of solitons \cite{Martin10a,Martin10b}. From another angle, an extension of the stochastic Bogoliubov approach treated each realization in the c-field ensemble as a source condensate to simulate pair scattering \cite{Kheruntsyan12,Lewis-Swan14}. An approach built from the SGPE-precursor master equation is hoped to alleviate some of the undesirable features of those approaches and to work even at temperatures for which density fluctuations are non-negligible. For example, stochastic Bogoliubov has spurious stimulated scattering into the quantum field where it overlaps with the c-field, properly treating only the high-energy modes \cite{Deuar11}, which restrict its application to the description of particles scattered there, such as in supersonic processes \cite{Deuar07,Kheruntsyan12,Lewis-Swan14}. Our approach here should be able to instead treat the quantum fluctuations in the complementary low-energy region ruined by stochastic Bogoliubov, where antibunching, quantum depletion, or a dynamical Casimir effect \cite{Jaskula12} can occur. In truncated Wigner, on the other hand, the virtual vacuum noise introduced into the c-field to emulate spontaneous scattering is not distinguished from the real particles. This led to spurious scattering of the vacuum, and e.g. produces an effectively negative occupation in high-energy modes \cite{Sinatra02,Deuar07}. 

We will first outline the SGPE method in Section~\ref{SGPE} along with showing its predictions for phase and density correlations in Sec.~\ref{SGPEFLUCT} for later comparison. Subsequently, the PPR treatment of the master equation is derived in Sec.~\ref{PPR}. Then, as a test case, we compare their predictions for the dynamics of a one-dimensional trapped quasicondensate after a quench of the interaction strength in Sec.~\ref{RES}. Quantum fluctuations are seen to cause the emergence of pairing from the initial thermal state. Density correlation waves appear similar to those predicted for a zero temperature quench, and they are not readily degraded by the thermal component. We also observe an additional reduction of phase coherence. The onset of quantum fluctuation effects is related to a breaking of the usual $gN$ invariance seen in c-field methods, which we will describe in Sec.~\ref{SCALE} and show its effects in Sec.~\ref{RES-g}.

%%%%%%%%%%%%%%%%%%%%%%%%%%%%%%%%%%%%%%%%%%%%%%%%%%%%%%%%%%%%%%%%%%%%%%%
\section{The SGPE method}
%%%%%%%%%%%%%%%%%%%%%%%%%%%%%%%%%%%%%%%%%%%%%%%%%%%%%%%%%%%%%%%%%%%%%%%
\label{SGPE}

\subsection{Summary of the method}

A feature of  c-field approaches in general  is a separation of the system into highly and lowly occupied modes, after which a detailed treatment is continued only for the highly occupied (low-energy) modes that are approximated by an ensemble of complex field amplitudes. The SGPE treats the effect of the remaining (high-energy) modes as a thermal and particle bath for the c-field. Such an approach can be contrasted to projected classical field methods such as the Projected Gross-Pitaevskii equation (PGPE) \cite{Davis01,Simula06} that remove the direct influence of the  high energy modes completely. 
The  derivation of the SGPE can be found in Refs. \cite{Stoof99,Stoof01,Gardiner03,Bradley05,Proukakis08}.  Its relationship to other c-field methods has been reviewed by Proukakis and Jackson \cite{Proukakis08}, and the method has been benchmarked in detail in recent works \cite{Cockburn11a,Gallucci12,Cockburn12,Bradley15} and extended to multicomponent gases \cite{Bradley14}. Some formulations explicitly include a projection of the c-field evolution onto the chosen low-energy subspace at each time-step \cite{Stoof99,Gardiner03,Rooney12,Rooney14}, which has been termed the SPGPE (stochastic {\it projected} GPE). We will base what follows in Sec.~\ref{PPR} on the derivation of Gardiner and Davis\cite{Gardiner03}, which is of this kind. 

The SGPE methods treat the system from a dynamical viewpoint, describing its state at nonzero temperature as an ensemble of complex wavefunctions $\phi(\bo{x},t)$. 
In a nutshell, the derivation proceeds as follows: the system is divided into two subsystems. One of them is represented by the field $\op{\phi}(\bo{x},t)$ and describes the low-lying modes of the ultracold gas. The second one is a thermal cloud of atoms whose energies are well above the typical energy of the condensate and its excitations~\cite{Stoof99}. 

Using a Hartree-Fock-like ansatz for the probability distribution of system states leads to separate probability distributions for high- and low-energy modes. 
By integrating over the low-energy modes, one finds that the thermal cloud may be treated by a quantum Boltzmann equation \cite{Stoof99,Gardiner03}. 
Integrating instead over the high-energy thermal cloud modes can be shown to lead to a master equation for the dynamics of the density matrix $\op{\rho}_C$ for the low-energy field $\op{\phi}$. 

\subsection{Master equation}

For later re-use in Sec.~\ref{PPR}, it is useful to present it here. Firstly, the low-energy subspace is spanned by the set of low-energy single-particle basis states $\{\,\ket{\psi_n}\,\}$, with normalized wavefunctions $\psi_n(\bo{x})$, so that a projector onto this operator subspace can be defined in the following way:
\begin{equation}\label{PC1}
\mc{P}_C = \sum_n \ket{\psi_n}\bra{\psi_n},
\end{equation}
while, correspondingly, for a spatial field $f(\bo{x})$,
\be\label{PC}
\mc{P}_Cf(\bo{x}) = \int d\bo{x}' \sum_n \psi_n(\bo{x})\psi_n^*(\bo{x}')f(\bo{x}')
\ee
and one defines
\begin{equation}\label{phi}
\op{\phi}(\bo{x}) = \mc{P}_C\op{\Psi}(\bo{x})
\end{equation}
in terms of the full Bose field $\op{\Psi}(\bo{x})$. 
Then, under appropriate conditions, the master equation for $\op{\rho}_C$ takes the form:

\begin{strip}\begin{equation}\label{master}
\frac{\partial\op{\rho}_C}{\partial t} = -\frac{i}{\hbar}\left[\op{H}_C,\op{\rho}_C\right] 
+ \int\!d\bo{x}\,\wb{G}(\bo{x})\left\{\left[\dagop{\phi}(\bo{x})\op{\rho}_C,\ \op{\phi}(\bo{x})\right] 
+ \left[\left(1-\frac{\mu}{k_BT}\right)\op{\phi}(\bo{x})\op{\rho}_C+\frac{\hbar}{k_BT}[\op{L}_C\op{\phi}(\bo{x})]\op{\rho}_C,\ \dagop{\phi}(\bo{x})\right]\right\} + \text{h.c.}\ 
\end{equation}\end{strip}

\noindent Here
\begin{equation}\label{HC}
\op{H}_C = \int\!d\bo{x}\,\dagop{\phi}(\bo{x})\left[H_{\rm sp}(\bo{x}) + \frac{g}{2}\,\dagop{\phi}(\bo{x})\op{\phi}(\bo{x})\right]\op{\phi}(\bo{x}),
\end{equation}
which includes the single-particle Hamiltonian density 
\begin{equation}\label{Hsp}
H_{\rm sp}(\bo{x}) = -\frac{\hbar^2}{2m}\nabla^2 + V(\bo{x})
\end{equation}
in an external potential $V(\bo{x})$. The contact inter-particle interactions have strength $g$, and $\mu$ is the chemical potential.
The growth/decay rate $\wb{G}(\bo{x})$ of the low-energy field can, in general, be spatially dependent. Finally, the low-energy frequency operator is
\begin{equation}\label{HGP}
\op{L}_C \op{\phi}(\bo{x}) = \mc{P}_C \left[ H_{\rm sp}(\bo{x}) \op{\phi}(\bo{x}) + g\,\dagop{\phi}(\bo{x})\op{\phi}(\bo{x})\op{\phi}(\bo{x})\right]. 
\end{equation}
All this corresponds to Eqs.~(83), (76), and (37) of \cite{Gardiner03}. The conditions imposed to obtain the above include: 
(i) disregarding the terms corresponding to scattering between condensate and thermal cloud atoms (in this context, see \cite{Rooney12}), as well as (ii) the usually small repulsive potential for the low energy field $\phi$ that comes from the thermal cloud, and (iii) assuming a sufficiently high thermal cloud temperature \cite{Gardiner03} that the c-field gain and decay rates ($G^{(+)}$ and $G^{(-)}$ in \cite{Gardiner03}, respectively) differ only by a relatively small amount as per (82) in \cite{Gardiner03}.  

\subsection{SGPE equation}

Following \cite{Gardiner03} and now treating the $\op{\phi}$ field in the truncated Wigner representation, with some auxiliary assumptions regarding the discarding of high-order terms, leads to 
the following nonlinear Langevin equation for samples $\phi(\bo{x})$ of a c-field ensemble:
\bea
\label{PSGPE}
\lefteqn{i \hbar \frac{\partial \phi}{\partial t} =} &&\\&& \mc{P}_C\left[(1-i \gamma) \left(H_{\rm sp} -\mu + g|\phi|^2 \right)\phi +  \sqrt{2\hbar \gamma k_BT}\, \eta\right].\nonu
\eea
Here, 
\be\label{gamma}
\gamma(\bo{x})= \frac{\hbar\,\wb{G}(\bo{x})}{k_BT}
\ee is a dimensionless decay rate that represents the coupling strength to the thermal bath. It can be spatially-varying, but is usually in practice taken small and constant, when equilibrium ensembles are desired. The $\eta$ are delta-correlated complex Gaussian stochastic noise fields, with the variances
\be
\langle \eta^{*}(\bo{x},t) \eta(\bo{x}^{\prime},t^{\prime}) \rangle=\delta (\bo{x}-\bo{x}^{\prime}) \delta(t-t^{\prime}).
\label{etadef}
\ee
 In practice they are approximated by a pair of real Gaussian random variables of variance $1/(2\Delta t \Delta x^d)$ in the real and imaginary directions that are independent at each point in time and $d$-dimensional space discretized with time steps $\Delta t$ and volume elements $\Delta x^d$. 
Thus, the effect of the high energy modes is described by an effective temperature $T$, chemical potential $\mu$, and the bath coupling strength $\gamma$.

With such c-field methods, one must separate out the low energy subspace that is to be treated using the field $\phi(\bo{x})$, a matter that has been studied in some detail \cite{Sinatra02,Brewczyk04,Brewczyk07,Blakie08,Witkowska09,Cockburn11a,Pietraszewicz15}.
It is common to make the simplest kind of split between low and high-energy modes, taking the low-energy subspace to be all plane-wave modes below a certain momentum cutoff $k_{\rm max} = \pi/\Delta x$, and this is what we will also do in this article. In that case, the projection in (\ref{PSGPE}) can be removed in the understanding that one works on a discretized numerical lattice in space, and that the upper half of the allowed momentum modes do not significantly contribute to the physics so that aliasing of the nonlinearity can be ignored. One then has the familiar form of the SGPE:
\be
\label{SGP_eq}
i \hbar \frac{\partial \phi}{\partial t} = (1-i \gamma) \left(H_{\rm sp} -\mu + g|\phi|^2 \right)\phi +  \sqrt{2\hbar \gamma k_BT}\, \eta.\ 
\ee

This equation is commonly used to obtain equilibrium states by evolving the ensemble  from essentially arbitrary starting states (e.g. vacuum) to long times, when the distribution stabilizes and becomes ergodic. 
The equilibrium particle number and energy are determined by the bath parameters $T$ and $\mu$, while $\gamma$ affects the time needed to reach equilibrium. 
One is able to obtain good results for temperatures in the quasicondensate or above-quasicondensate regimes, where the thermal fluctuations in both density and phase can be much higher than for the Bogoliubov description, and the condensate fraction can be small \cite{Cockburn11a,Gallucci12}.

\begin{figure}[ht!]
\begin{center}
\includegraphics[width=\columnwidth]{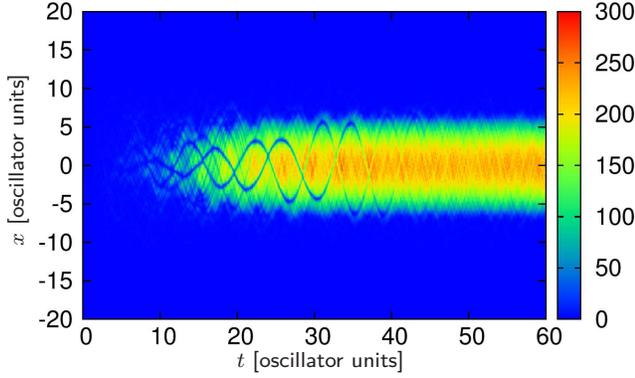}
\end{center}\vspace*{-4mm}
\caption[]{ The generation of a single sample $\phi(x)$ of the thermal equilibrium ensemble for a harmonically trapped 1D Bose gas. Color shows the local density $n(x,t)=|\phi(x,t)|^2$ of the gas during its time evolution under (\ref{SGP_eq}). All quantities are in trap harmonic oscillator units where $\hbar=m=a_{\rm ho}=\sqrt{\hbar/2m\omega}=1$. Thermal cloud bath parameters are $T=13.89$, $\mu=22.41$, $g=0.1$, and $\gamma=0.01$. 
These parameters correspond to the coldest of the cases studied in detail in Secs.~\ref{SGPEFLUCT} and~\ref{RES-T}, and are fairly close in properties to the trapped gas of the experiment of Ref.~\cite{Bucker11}. 
\label{fig:dens_SGPE}}
\end{figure}

The time evolution of such a calculation is shown in figure~\ref{fig:dens_SGPE} for a trapped 1D Bose gas. 
The simulation starts from a vacuum initial condition $\phi(x,0)=0$, and evolves to an equilibrium trapped quasicondensate. Ensemble properties of such growth were considered in detail in \cite{Cockburn09,Cockburn11a}. The figure here shows a single realization of a wavefunction in the ensemble. One notable feature is the spontaneous appearance of two deep solitons in the gas, and their later disappearance as an equilibrium quasicondensate is reached. Such effects have been seen previously during the evaporative cooling and subsequent thermalization of a 1D gas \cite{Witkowska11,Schmidt12}, or other sudden disturbances \cite{ChaoFeng,Hamner11,Liu14}.

%%%%%%%%%%%%%%%%%%%%%%%%%%%%%%%%%%%%%%%%%%%%%%%%%%%%%%%%%%%%%%%%%%%%%%%
\subsection{Fluctuations in the SGPE}
%%%%%%%%%%%%%%%%%%%%%%%%%%%%%%%%%%%%%%%%%%%%%%%%%%%%%%%%%%%%%%%%%%%%%%%
\label{SGPEFLUCT}

Let us consider now the predictions generated by the SGPE for density and phase fluctuations in a quasicondensate, for comparison with the fuller equation derived in Sec.~\ref{PPR}. 
Similarly to  figure~\ref{fig:dens_SGPE}, we take the following parameters, chosen to match earlier benchmarking studies of trapped 1D gases \cite{Cockburn11a}:   
In harmonic oscillator units ($\hbar=m=a_{\rm ho}=\sqrt{\hbar/m\omega}$) for a 1D trap of angular frequency $\omega$, we take an interaction strength of $g=0.1$. 
The thermal cloud bath parameters are $\mu=22.41$, $\gamma=0.01$, and we will use three temperatures: $T= 0.62\mu,\ 1.24\mu,\ 1.91\mu$, which can be compared to the characteristic phase coherence temperature \cite{Petrov00}
\be\label{Tphi}
T_{\phi}\approx N\,\frac{(\hbar\omega)^2}{\mu} \approx \frac{4\sqrt{2\mu}}{3g}\,\frac{\hbar^2\omega}{\sqrt{m}},
\ee
which is  $T_{\phi}=3.98\mu$ in our case. So, we have $T=$ 0.156, 0.311 and 0.480 $T_{\phi}$ here. The trapped ideal gas critical temperature is $T_c\approx N/\log 2N = 10.76\mu$ \cite{Ketterle96}. 

These parameters are like those used in the study  \cite{Cockburn11a} apart from a simple variable change in the SGPE (discussed in Sec.~\ref{SCALE}) that leads to a  $10 \times$ increase in $g$. 
The reason for the scaling is to be closer to experimental values, something that will become relevant once quantum fluctuations are added in Sec.~\ref{RES}, breaking the SGPE scaling. 
For example, our parameters correspond to ${}^{87}{\rm Rb}$ atoms in a trap with frequencies $\nu$ of $520 \times 520 \times 30.2$ Hz, at temperatures of 20, 40, and 62 nK, respectively, which we will call our ``reference system''. The number of atoms is $\approx2000$. 
This case can be compared to a recent experiment in Vienna \cite{Bucker11}, that had about 700 atoms at 40 nK, with a slightly more elongated trap of axial frequency 16.3 Hz. 

To generate the thermal equilibrium state, simulations start in vacuum, and continue for a time of $60\hbar\omega$, which appears sufficient for equilibration of a single realization (see figure~\ref{fig:dens_SGPE}). We use 10 000 realizations to reduce noise in the density correlations.

\begin{figure}[htb]
\begin{center}
\includegraphics[width=0.75\columnwidth]{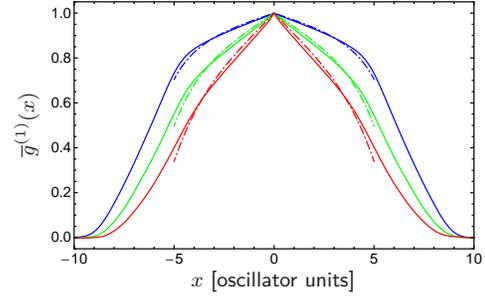}
\end{center}\vspace*{-4mm}
\caption{ The phase correlation function $\wb{g}^{(1)}(x)$ when the system with $\mu=22.41$ is described by the SGPE. $T=0.156T_{\phi}$ - blue, $T=0.311T_{\phi}$ - green, $T=0.480T_{\phi}$ - red. 
Note the expected increasing reduction of coherence length as $T$ grows. 
Dot-dashed lines 
show thermal quasicondensate estimates (\ref{Thqcg1}) with an effective $\mu_{\rm eff}=\mu-x^2/2$.
\label{fig:g1_SGP_u}}
\end{figure}

We concentrate on correlation functions in the center of the cloud or in momentum-space. 
The two-point normalized correlation function 
\bea 
g^{(1)}(x_1,x_2) = \frac{\langle \dagop{\Psi}(x_1) \op{\Psi}(x_2)\rangle}{\sqrt{n(x_1)\,n(x_2)}}\! \to\! \frac{\langle \phi(x_1)^*\phi(x_2)\rangle_{\rm ens\!\!\!\!\!}}{\sqrt{n(x_1)\,n(x_2)}}\qquad\mbox{}
\label{g1}
\eea
describes the phase coherence, with normalization by the local density $n(x)=\langle\dagop{\Psi}(x)\op{\Psi}(x)\rangle \to \langle|\phi(x)|^2\rangle_{\rm ens}$. 
The right-hand expressions indicated with ``$\to$'', are the averages to be carried out over the statistical ensemble of samples generated by  the SGPE. Their precision increases with the size of the ensemble. 
To obtain a better signal-to-noise ratio for the spatial correlations in the center of the trap, the correlations were locally averaged over starting points $x'$ lying in the center 30\% of the cloud ($|x'|<x_c=2$), as per
\bea 
\wb{g}^{(n)}(x) = \frac{1}{2x_c} \int_{|x'|<x_c} dx'\ g^{(n)}(x',x'+x). 
\label{g12_usr}
\eea
where $n=1$ or $2$.
The phase correlations for the reference system are shown in figure~\ref{fig:g1_SGP_u}. There is a linear loss of phase coherence with distance and temperature, which is expected for a quasicondensate whose phase fluctuations are dominated by thermal effects. In that case the decay of phase coherence can be estimated \cite{Petrov00,Mora03} as:
\be
g^{(1)}(x) \approx e^{-x/L_{\phi}}, \quad\text{ where }\quad L_{\phi} = \frac{2\mu}{gT}\left(\frac{\hbar^2}{mk_B}\right).
\label{Thqcg1}
\ee
A first easy correction can be obtained by taking a local effective chemical potential $\mu_{\rm eff}(x) = \mu-\frac{x^2}{2}$, leading to local $L_{\phi}(x)$. Such an estimate is shown for comparison in figure~\ref{fig:g1_SGP_u} as dot-dashed lines. The match is quite good until trap edge effects kick in at $|x|\approx5$.
Phase correlations in similar regimes have been investigated e.g. in \cite{Cockburn09,Cockburn11a} and compared to experiment \cite{Gallucci12}. 

\begin{figure}[htb]
\begin{center}
\includegraphics[width=0.75\columnwidth]{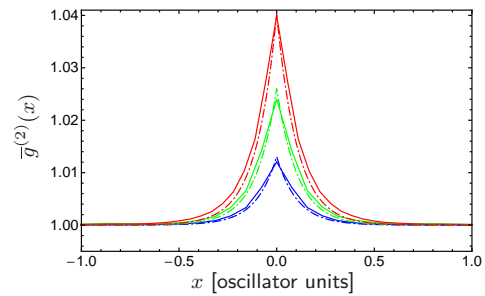}
\end{center}\vspace*{-4mm}
\caption{ The density correlation function $\wb{g}^{(2)}(x)$  when the system with $\mu=22.41$ is described by the SGPE. $T=0.156T_{\phi}$ - blue, $T=0.311T_{\phi}$ - green, $T=0.480T_{\phi}$ - red.
Note the expected growth of bunching with $T$.
Dot-dashed lines 
show thermal quasicondensate estimates (\ref{Thqcg2}).
\label{fig:g2_SGP_u}}
\end{figure}

The second-order correlation function 
\bea 
g^{(2)}(x_1,x_2)&=&\frac{\langle \dagop{\Psi}(x_1) \dagop{\Psi}(x_2) \op{\Psi}(x_1) \op{\Psi}(x_2)\rangle}{n(x_1)n(x_2)}\nonumber\\&& \to \frac{\langle |\phi(x_1)|^2|\phi(x_2)|^2 \rangle_{\rm ens}}{n(x_1)n(x_2)}
\label{g2}
\eea
describes the density fluctuations, and is shown in figure~\ref{fig:g2_SGP_u}. Here one sees weak bunching, growing with temperature, as expected in a thermal quasicondensate.
For a quasicondensate in the thermal regime \cite{Deuar09}, the estimate for a uniform gas with density $n$ is:
\be
g^{(2)}(x) \approx 1+\frac{T}{n^{3/2}\sqrt{g}}\left(\frac{k_B\sqrt{m}}{\hbar}\right)e^{-2x/\xi_{\rm heal}}
\label{Thqcg2}
\ee
where 
\be\label{heall}
\xi_{\rm heal}=\hbar/\sqrt{mgn}
\ee
is the healing length. Taking the Thomas-Fermi estimate of density in the center of the trap, $n\to n_0=\mu/g$, one obtains the estimates shown for comparison in figure~\ref{fig:g2_SGP_u} as dot-dashed lines. These  match very well. 

However, it is also known that for low enough temperatures, the uniform gas displays antibunching, i.e. $g^{(2)}(0)<1$, an effect that is caused by two-body repulsion, and not treated by c-field descriptions. For a dilute zero temperature gas, $g^{(2)}(0) \approx 1 - 2\sqrt{g}/\pi\sqrt{n}$ \cite{Kheruntsyan03}. An exact calculation from the Yang-Yang exact solution for the uniform gas \cite{Yang69}
using the central density estimate  $n_0=\mu/g=224.1$, 
gives the following values for the three increasing temperatures used here: $g^{(2)}(0) = 0.98986, 0.99604$, and $1.0042$.
This does not match the SGPE result, with a particularly glaring discrepancy at the lowest temperature, where one has anti-bunching in the true gas instead of bunching. 

\begin{figure}
\begin{center}
\includegraphics[width=0.75\columnwidth]{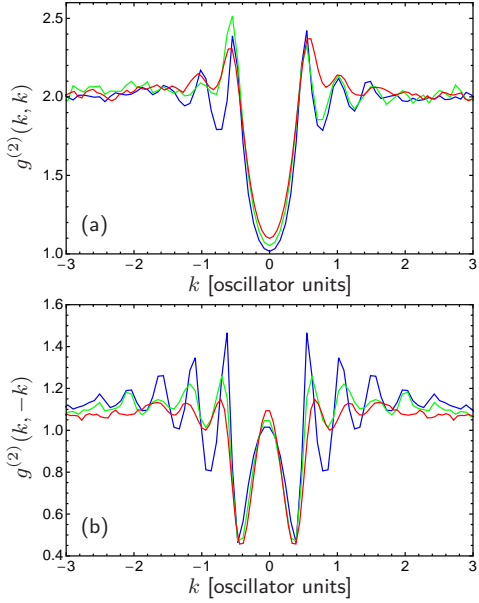}
\end{center}\vspace*{-4mm}
\caption[]{ 
Normalized density fluctuations in momentum space $g^{(2)}(k,k)$ (panel a) and the counter-propagating pair correlation function $g^{(2)}(k,-k)$ (panel b). SGPE calculation with $\mu=22.41$ and $T=0.156T_{\phi}$ - blue, $T=0.311T_{\phi}$ - green, $T=0.480T_{\phi}$ - red. Statistical uncertainty is $\sim\pm0.05$. 
\label{fig:g2kk_SGP}}
\end{figure}

In k-space, an analogous expression to (\ref{g2}) holds:
\bea 
g^{(2)}(k,k')=\frac{\langle \dagop{\Psi}(k) \dagop{\Psi}(k') \op{\Psi}(k) \op{\Psi}(k')\rangle}{n(k)n(k')}.
\label{g2kk}
\eea
This gives information about thermal excitations and atom pairing in the system. Thermally occupied modes have Hanbury Brown-Twiss-like (HBT) density fluctuations: $g^{(2)}(k,k) = 2$, while pairing between counter-propagating atoms  would be evidenced by increased density correlations between them: $g^{(2)}(k,-k) >1$. These two quantities are shown in figure~\ref{fig:g2kk_SGP}.

\begin{figure}
\begin{center}
\includegraphics[width=0.75\columnwidth]{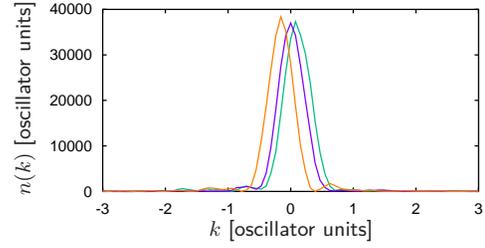}
\end{center}\vspace*{-4mm}
\caption[]{ 
Momentum density $n(k)$ for several realizations when the system with $g=0.1$, $\mu=22.41$, and $T=0.156T_{\phi}$ is described by the SGPE. 
\label{fig:nk_SGP}}
\end{figure}

For comparison, typical momentum densities are shown in figure~\ref{fig:nk_SGP}. 
By comparing figures, one sees that for momenta well beyond the main cloud there is the expected HBT behavior and no pairing.
The main features seen at low k have rather trivial causes, but require some explanation. In the presence of both condensate and excitations two effects modify the simplest picture with $g^{(2)}(k,k') = 1$ in the condensate and thermal  $g^{(2)}(k,k)=2$ beyond. 

First -- thermal fraction. Consider a toy model where the wavefunction at a given momentum $k$, $\phi(k)=\phi_0(k)+\varepsilon(k)$ consists of a condensate fraction $n_0(k)\le1$ in wavefunction $\phi_0(k)$ and independent Gaussian fluctuations $\varepsilon(k)$, such that its ensemble averages are $\langle\varepsilon\rangle=0$, and $\langle|\varepsilon|^4\rangle=2\langle|\varepsilon|^2\rangle^2$. Then it is easily shown that 
\be
g^{(2)}(k,k) = 2-n_0(k)^2. 
\ee
This accounts for the bulk of the variation in figure~\ref{fig:g2kk_SGP}(a).

Secondly -- center-of-mass motion. Inspection of single realizations of $\phi(x)$ in the SGPE ensemble reveals that they are typically somewhat narrower than the ensemble mean, as seen by the relative displacement of individual realizations in figure~\ref{fig:nk_SGP}. This is due to appreciable center-of mass displacements in the trap. Consider then another toy model, when the wavefunction in individual realizations is a randomly displaced condensate $\phi(k)= \phi_0(k+\delta)$, with the displacement $\delta$ Gaussian distributed with standard deviation $\sigma$: $P(\sigma)\propto e^{-\delta^2/2\sigma^2}$. The mean density is then $\langle|\phi(k)|^2\rangle = \int d\delta P(\delta) n_0(k+\delta)$ in terms of the un-displaced condensate density $n_0(k)=|\phi_0(k)|^2$. A Taylor series expansion in small $\delta$ then gives
$\langle|\phi(k)|^2\rangle\approx n_0(k) + \frac{1}{2}\sigma^2 (\partial^2n_0(k)/\partial k^2)$. 
A similar calculation provides an expression for $\langle|\phi(k)|^2|\phi(\pm k)|^2\rangle\approx n_0(k)^2+\sigma^2[(\partial n_0(k)/\partial k)^2\pm n_0(k)(\partial^2n_0(k)/\partial k^2)]$, leading to a final estimate of the pair correlation function (when $\sigma$ is small) as:
\be
g^{(2)}(k,\pm k) \approx 1 \pm \sigma^2 \left(\frac{1}{n_0(k)}\frac{\partial n_0(k)}{\partial k}\right)^2.
\ee 
From this, one can see how the apparent value of $g^{(2)}(k,-k)$ can be lowered below 1 for counter-propagating atoms, and raised above the otherwise expected value of 2 for $g^{(2)}(k,k)$ correlations by the rather trivial center-of-mass motions.
In particular, the effect is most pronounced at the edge of the condensates where the ratio of gradient to density is highest. This explains the form of the excursions below unity in figure~\ref{fig:g2kk_SGP}(b) and above two in figure~\ref{fig:g2kk_SGP}(a). 
From equipartition arguments, the center-of mass energy per particle is $T/\wb{N}$ on average, which corresponds to a typical COM momentum of $k_{COM}\sim0.1$ in the example system treated here. In comparison, the condensate width in momentum space is approximately the inverse of the Thomas-Fermi radius, i.e. $\sim1/\sqrt{2\mu}\approx0.15$. Taken together, these values validate that spontaneous center-of-mass motion may be significant for this system.

Finally, regarding pairing, consider a state that is close to being a condensate,  such that an expansion of the Bose field into a dominant wavefunction $\phi_0(x)$ and relatively small fluctuations $\delta\op{\Psi}(x)$ as per the Bogoliubov approach is reasonable. That is, $\op{\Psi}(x) = \phi_0(x) + \delta\op{\Psi}(x)$. An expansion of the interaction term in the Hamiltonian to lowest relevant order in the fluctuations gives both potential terms of the form  $g|\phi_0(x)|^2\delta\dagop{\Psi}(x)\delta\op{\Psi}(x)$, and pair production terms of the form $g\phi_0(x)^2\delta\dagop{\Psi}(x)^2$. The latter should lead to the appearance of some level of pairing between counter-propagating atoms in the system. 
The lack of such a clear pairing signature in figure~\ref{fig:g2kk_SGP}(b) is something that we expect a fuller theory than the SGPE to rectify.

%%%%%%%%%%%%%%%%%%%%%%%%%%%%%%%%%%%%%%%%%%%%%%%%%%%%%%%%%%%%%%%%%%%%%%%
\section{Positive-P representation}
%%%%%%%%%%%%%%%%%%%%%%%%%%%%%%%%%%%%%%%%%%%%%%%%%%%%%%%%%%%%%%%%%%%%%%%
\label{PPR}

Let us treat the master equation (\ref{master}) from which the SGPE originates using the exact mapping to a positive-P representation (PPR) instead of the usual truncated Wigner approximation. 

\subsection{Formalism}

The PPR is an expansion of the density operator in terms of an off-diagonal coherent state projector basis  $\op{\Lambda}$. For a single mode it is:  
\be
\op{\rho} = \int P(\alpha,\wt{\alpha}) \op{\Lambda} (\alpha,\wt{\alpha})\,d^2\alpha\,d^2\wt{\alpha}, 
\label{dens_pp}
\ee
where \mbox{$\op{\Lambda} =|\alpha\rangle \langle\wt{\alpha}^*|\,/\,\langle\wt{\alpha}^*|\alpha \rangle$} with bosonic coherent states \mbox{$\ket{\alpha}=\exp(\alpha\,\dagop{a}-|\alpha|^2/2)\ket{0}$}  having phase $\angle\alpha$ and mean particle number $|\alpha|^2$. The distribution function in the phase space spanned by the ``bra'' and ``ket'' amplitudes $\{\alpha,\wt{\alpha}\}$ is $P(\alpha,\wt{\alpha})$ and can be chosen such that it remains real and positive \cite{Drummond80}.

The underlying idea here is that this is targeted towards expressing the many-body state of a quantum system as a distribution over simpler basis states that are local to each mode. For large systems such as we are interested in here, the aim is to interpret the positive real distribution $P$  as a probability of the basis states, or ``realizations'' of the system, and sample them stochastically. This enormously reduces the size of the description of the system, down to an ensemble of realizations, at the cost of introducing statistical uncertainty. 

The definition (\ref{dens_pp}) can be extended straightforwardly to a many-mode system, such as the set of basis states $\ket{\psi_n}$ in our low-energy subspace as per (\ref{PC1}). 
The many-mode operator basis $\op{\Lambda}$ is taken to just be an operator product of the local operators:
\begin{equation}\label{ppmm}
\op{\Lambda} = \bigotimes_{n} \op{\Lambda}_n(\alpha_n,\wt{\alpha}_n) = \bigotimes_{n} \frac{\ket{\alpha_n}\bra{\wt{\alpha}_n^*}}{\braket{\wt{\alpha}^*_n}{\alpha_n}}.
\end{equation}
with coherent state amplitudes $\alpha_n$ and $\wt{\alpha}_n$ for each mode. Since the boson wavefunction in the low energy subspace can be expanded as
\be\label{expandphi}
\op{\phi}(\bo{x}) = \sum_n \psi_n(\bo{x})\, \op{a}_n,
\ee
then two corresponding ``bra'' and ''ket'' c-fields can be constructed from the basis-state amplitudes:
\numparts
\bea
\phi(\bo{x}) &=& \sum_n\psi_n(\bo{x})\,\alpha_n \nonu\\
\wt{\phi}(\bo{x}) &=& \sum_n\psi_n(\bo{x})\,\wt{\alpha}_n, \label{phifield}
\eea\endnumparts
So that we can also write $\op{\Lambda}(\phi(\bo{x}),\wt{\phi}(\bo{x}))$. In this way, $\op{\Lambda}$ is an off-diagonal projector between coherent states in the $\phi(\bo{x})$ and $\wt{\phi}(\bo{x})$ orbitals, with mean occupations of  $\int d\bo{x}\, |\phi(\bo{x})|^2$ and $\int d\bo{x}\, |\wt{\phi}(\bo{x})|^2$, respectively, and the distribution $P$ is over all possible pairs of spatial wavefunctions $\phi(\bo{x})$ and $\wt{\phi}(\bo{x})$ in the subspace projected onto by $\mc{P}_C$.

The final aim is to map the master equation (\ref{master}) for $\op{\rho}_C = \int P \op{\Lambda}(\phi(\bo{x}),\wt{\phi}(\bo{x}))\,\mc{D}\phi(\bo{x})\,\mc{D}\wt{\phi}(\bo{x})$
into equations for the samples $\phi(\bo{x})$ and $\wt{\phi}(\bo{x})$. 
The usual procedure to do this \cite{Drummond80,WallsMilburn,QuantumNoise} uses the correspondence relations between local Bose field operators $\dagop{a}_n$ and derivatives:
\bea
\op{a}_n \op{\Lambda} &=& \alpha_n\, \op{\Lambda} \nonu\\
\dagop{a}_n \op{\Lambda} &=& \left( \wt{\alpha}_n^* + \frac{\partial}{\partial \alpha_n} \right) \op{\Lambda} \nonu\\
\op{\Lambda} \dagop{a}_n &=& \wt{\alpha}_n^* \op{\Lambda} \label{corresp}\\
\op{\Lambda}\op{a}_n &=& \left( \alpha_n + \frac{\partial}{\partial \wt{\alpha}_n^*} \right) \op{\Lambda} 
\eea
to derive a Fokker-Planck equation, which in general takes the form
\be
\frac{\partial P(\vec{v})}{\partial t} = \Big\{-\frac{\partial}{\partial v_{\mu}}A_{\mu}(\vec{v}) +\frac{1}{2}\frac{\partial}{\partial v_{\mu}}\frac{\partial}{\partial v_{\nu}}D_{\mu \nu} (\vec{v})\Big\}P(\vec{v}),\ 
\label{FPE-general}
\ee
where $\mu,\nu$ label the phase-space variables $v_{\mu}$ that can be any of the $\alpha_n$ or $\wt{\alpha}_n$.  $A^{\mu}$ is the drift vector and $D^{\mu \nu}$ is the diffusion matrix, which can in general depend on all the variables $\vec{v}=\{\dots,v_{\mu},\dots\}$. 
The Fokker-Planck equation can then be mapped onto a set of coupled, complex Ito stochastic differential equations:
\be
\frac{dv_{\mu}}{dt} = A_{\mu}(\vec{v}) + \sum_{\nu} B_{\mu \nu}(\vec{v})\zeta_{\nu}(t). 
\ee
where the noise matrix $B$ satisfies the matrix equation $D=BB^T$, and $\zeta_{\nu}(t)$ are delta-time-correlated, independent, real white noise fields with variance 
\be
\langle\zeta_{\mu}(t)\zeta_{\nu}(t')\rangle = \delta_{\mu\nu}\delta(t-t').
\ee

\subsection{Low-energy PPR equations}

For the master equation (\ref{master}), one obtains an exact mapping to the following Fokker-Planck equation:

\begin{strip}
\bea\label{FPE}
\hbar\frac{\partial P}{\partial t} &=& \quad
\sum_n\frac{\partial}{\partial\alpha_n}\int\!d\bo{x}\,\psi_n^*(\bo{x})     \left\{\left(i+\gamma(\bo{x})\right) \left[H_{\rm sp}+g\,\wt{\phi}^*(\bo{x})\phi(\bo{x})\right]-\gamma(\bo{x})\mu\right\}\phi(\bo{x}) P\\
&&+\sum_n\frac{\partial}{\partial\wt{\alpha}_n^*}\int\!d\bo{x}\,\psi_n(\bo{x})\left\{\left(-i+\gamma(\bo{x})\right)\left[H_{\rm sp}+g\,\phi^*(\bo{x})\wt{\phi}(\bo{x})\right]-\gamma(\bo{x})\mu\right\}\wt{\phi}^*(\bo{x}) P\nonu\\
&&-g\sum_{nm}\frac{\partial^2}{\partial\alpha_n\partial\alpha_m}\int\!d\bo{x}\left[\frac{i}{2}+\gamma(\bo{x})\right]\psi_n^*(\bo{x})\phi(\bo{x})^2\psi_m^*(\bo{x}) P\nonu\\
&&+g\sum_{nm}\frac{\partial^2}{\partial\wt{\alpha}_n^*\partial\wt{\alpha}_m^*}\int\!d\bo{x}\left[\frac{i}{2}-\gamma(\bo{x})\right]\psi_n(\bo{x})\wt{\phi}^*(\bo{x})^2\psi_m(\bo{x}) P
+\sum_{nm} \frac{\partial^2}{\partial\alpha_n\partial\wt{\alpha}^*_m}\int\!d\bo{x}\,2\gamma(\bo{x})k_BT\psi_n^*(\bo{x})\psi_m(\bo{x})P\nonu
\eea
\end{strip}

\noindent
with the usual definition (\ref{gamma}) of $\gamma$.
Use was made of the orthogonality of the mode wavefunctions:
\be\label{orthnm}
\int\!d\bo{x}\,\psi_n^*(\bo{x})\psi_m(\bo{x}) = \delta_{nm}.
\ee

One obtains Langevin equations for the mode amplitudes, and then immediately for the c-fields via (\ref{phifield}), since the mode wavefunctions $\psi_n(\bo{x})$ are time-independent. 
It is also convenient to add a global phase evolution of $\mu t/\hbar$ to $\phi$ and $\wt{\phi}$. The equations, with $\bo{x}$ and $t$ dependence of all fields ($\phi, \wt{\phi}, V, \xi, \wt{\xi}, \gamma$) implied, are:

\begin{strip}
\bea
i \hbar \frac{d\phi}{dt}      &=& \mc{P}_C\left[(1-i\gamma)\left(-\frac{\hbar^2 \nabla^2}{2 m}+V -\mu + g\,\phi\, \wt{\phi}^{*}\right)\phi  +\sqrt{i \hbar g\,(1-2i\gamma)}\, \phi\,\xi
+\sqrt{2\hbar \gamma k_BT }\ \,\eta\right]\nonu\\
i \hbar \frac{d\wt{\phi}}{dt} &=& \mc{P}_C\left[(1-i\gamma)\left(-\frac{\hbar^2 \nabla^2}{2 m}+V -\mu + g\,\wt{\phi}\, \phi^{*}\right)\wt{\phi} +\sqrt{i \hbar g\,(1-2i\gamma)}\, \wt{\phi} \,\wt{\xi}
+\sqrt{2\hbar \gamma k_BT }\ \,\eta\right].\label{ppr-eq}
\eea
\end{strip}

\noindent
This explicitly includes projection onto the low energy subspace (\ref{PC}) at every time step.
The independent real white noise fields $\xi(\bo{x},t)$ and $\wt{\xi}(\bo{x},t)$ individually have variances 
\be
\langle\xi(\bo{x},t)\xi(\bo{x}',t')\rangle = \delta(\bo{x}-\bo{x}')\delta(t-t'). 
\label{xidef}
\ee
In practice, this is implemented with independent, real Gaussian noises at each numerical lattice point and time step $\Delta t$ that have a variance of $1/\Delta t\Delta x^d$.  The properties of $\eta(\bo{x},t)$ are given by (\ref{etadef}).

\subsection{Comparison with the SGPE}
\label{COMP}

The equations (\ref{ppr-eq}) are a generalization of the PSGPE of (\ref{PSGPE}) to include the full quantum mechanics of the $\op{\phi}$ low-energy field. There are three main differences: (i) The separation into ``bra'' and ``ket'' fields, (ii) the addition of the ``quantum noise'' stochastic terms with real noises $\xi$ and $\wt{\xi}$, and (iii) a replacement of $|\phi|^2$ with $\phi\wt{\phi}^*$ or its complex conjugate as estimators for the local density. 

The presence of the two fields $\phi$ and $\wt{\phi}$ allows for the incorporation of the nonzero commutation relation for the Bose field $\op{\phi}$, i.e. 
\be\label{commutation}
[\op{\phi}(\bo{x}),\dagop{\phi}(\bo{x}')] = \mc{P}_c(\bo{x},\bo{x}') = \sum_n \psi_n(\bo{x})\psi_n(\bo{x}')
\ee
 Expectation values of all quantum observables $\op{O}$  are calculated by the following procedure, which can be derived from the definition of the representation (\ref{dens_pp}) and the operator identities (\ref{corresp}) in a  straightforward way \cite{WallsMilburn}:
\begin{enumerate}
\item One first expresses the operator $\op{O}$ in its normally ordered form $:\op{O}:$  (i.e. by rearranging its expression with the help of (\ref{commutation}) so that all creation operators $\dagop{\phi}$ are to the left of all annihilation operators $\op{\phi}$  in all the terms).
\item A functional $f_O[\phi,\wt{\phi}]$ is obtained by replacing $\op{\phi}(\bo{x})\to\phi(\bo{x})$ and $\dagop{\phi}(\bo{x})\to\wt{\phi}(\bo{x})^*$ in\ $:\op{O}:$
\item The statistical mean of $f_O$, that is, $\langle {\rm Re}\left\{f_O[\phi,\wt{\phi}]\,\right\} \rangle_{\rm ens}$ converges to the quantum mechanical average $\langle\op{O}\rangle$ as the size of the statistical ensemble grows. 
\end{enumerate}
For example, the one-body density matrix is evaluated as
\be\label{rho1}
\rho_1(\bo{x},\bo{x}') = \langle{\rm Re}\left[\wt{\phi}(\bo{x})^*\phi(\bo{x}')\right]\rangle_{\rm ens}
\ee
Note that the requirement that the functional $f_O[\phi,\wt{\phi}]$ is obtained from the normal-ordered form of the operator leads to effectively nonzero commutation relations. 
For example, $:\!\op{\phi}(\bo{x}')\dagop{\phi}(\bo{x})\!:\quad\! = \dagop{\phi}(\bo{x})\op{\phi}(\bo{x}') + \mc{P}_C(\bo{x},\bo{x}')$, so that the functional evaluated to calculate the expectation value of $\op{\phi}(\bo{x}')\dagop{\phi}(\bo{x})$ 
is greater by $\mc{P}_C(\bo{x},\bo{x}')\approx\delta^d(\bo{x}-\bo{x}')$ than that used to calculate the mean density,  $\dagop{\phi}(\bo{x})\op{\phi}(\bo{x}')$. This is as required by full quantum mechanics. 

When taking the plane-wave basis on lattice spacing $\Delta x$ as with the plain SGPE (\ref{SGP_eq}) we have

\begin{strip}
\bea
i \hbar \frac{d\phi}{dt}      &=& (1-i\gamma)\left(-\frac{\hbar^2 \nabla^2}{2 m}+V -\mu + g\,\phi \wt{\phi}^{*}\right)\phi  +\sqrt{i\hbar g(1-2i\gamma)}\,\phi\,\xi +\sqrt{2\hbar \gamma k_BT }\,\eta\nonu\\
i \hbar \frac{d\wt{\phi}}{dt} &=& (1-i\gamma)\left(-\frac{\hbar^2 \nabla^2}{2 m}+V -\mu + g\,\wt{\phi} \phi^{*}\right)\wt{\phi}+\sqrt{i\hbar g(1-2i\gamma)}\,\wt{\phi}\,\wt{\xi}+\sqrt{2\hbar \gamma k_BT }\,\eta.\label{PPR-eq}
\eea
\end{strip}

\noindent
These equations are very similar to those conjectured earlier by a heurstic approach \cite{Swislocki14}. The difference is a $\sqrt{1-2i\gamma}$ factor on the quantum noise instead of $(1-i\gamma)$. These become equal as $\gamma$ becomes small. 

\begin{figure}
\begin{center}
\includegraphics[width=\columnwidth]{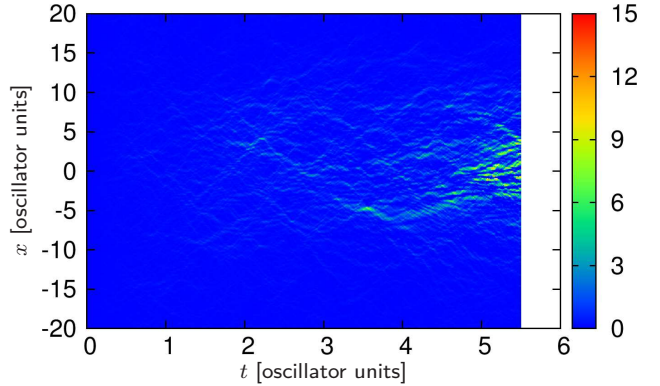}
\end{center}\vspace*{-4mm}
\caption[]{ An attempt to generate a sample  of the thermal equilibrium ensemble with the raw PPR equations (\ref{PPR-eq}). The density $n(x,t)={\rm Re}[\wt{\phi}^*\phi]$ calculated via (\ref{rho1}) is shown. All parameters like in figure~\ref{fig:dens_SGPE}, except for the markedly shorter timescale. The white space on the right indicates the onset of catastrophic noise amplification. 
\label{fig:dens_hybrid}}
\end{figure}

While the equations (\ref{ppr-eq}) and (\ref{PPR-eq}) incorporate the full quantum dynamics of the system, they also suffer from a serious problem if one is interested in long time scales.
The nonlinearity in the equations amplifies the fluctuations that are being input via $\xi(t)$ and $\wt{\xi}(t)$, which leads to unmanageable statistical error after some time $t_{\rm sim}$. 
An estimate for this time was obtained for systems with no thermal bath:\cite{Deuar06}  
\bea 
t_{\rm sim} \approx \frac{2\hbar(\Delta x)^{d/3}}{g(n_{\rm max})^{2/3}}
\label{sim_time}
\eea
where $n_{\rm max}$ is the maximum density in the system. 
While sufficiently strong dissipation is known to stabilize stochastic equations coming from the PPR \cite{Gilchrist97,Deuar06}, the required strength of $\gamma$ is larger than that found in our example calculations. 
Since reaching an equilibrium thermal state requires long time evolution, this usually precludes using the raw PPR equations Eqs. (\ref{PPR-eq}) for this purpose.
For example, growing the gas from vacuum with the equations (\ref{PPR-eq}) in the same manner as was done in  figure~\ref{fig:dens_SGPE} with the SGPE leads to what is shown in figure\ref{fig:dens_hybrid}. 

%%%%%%%%%%%%%%%%%%%%%%%%%%%%%%%%%%%%%%%%%%%%%%%%%%%%%%%%%%%
%\subsection{Interaction strength scaling}
\subsection{Onset of quantum fluctuations}
%%%%%%%%%%%%%%%%%%%%%%%%%%%%%%%%%%%%%%%%%%%%%%%%%%%%%%%%%%%
\label{SCALE}

A useful quantity to describe the quantum granularity, or degree to which a semiclassical description is inaccurate, is the Lieb-Liniger dimensionless interaction strength $\gamma_{LL}=m g / \hbar^2n$ introduced in \cite{Lieb63} for 1D. (It is not to be confused with the unrelated $\gamma$ bath coupling strength used in the stochastic equations). 

There is a continuous symmetry of the SGPE description that remains even after all quantities have been expressed in dimensionless units as
\be\label{SGP_uu}
i\frac{\partial\phi}{\partial t} = (1-i\gamma)(H_{\rm sp}-\mu+g|\phi|^2)\phi  + \sqrt{2\gamma T}\eta
\ee
 along with the normalization condition that the mean number of particles is $\wb{N} = \int d\bo{x}\,|\phi(\bo{x})|^2$.
Namely, the equation is unchanged under the following transformation with one real parameter, $\lambda>0$:
\bea
g &\to& \lambda\, g \nonu\\
\phi(\bo{x}) &\to& \phi(\bo{x})/\sqrt{\lambda} \label{scaling}\\
T &\to& T/\lambda\nonu
\eea
while $\wb{N} \to \wb{N}/\lambda$. Since there is no scaling of position or time coordinates (nor of $\mu$, $V$ or $\gamma$), this property remains true also when the system is discretized onto a numerical lattice.
Note though, that taking into account the physics of the problem in a way that goes beyond the equation itself, the most appropriate cutoff $k_{\rm max}$ is not generally invariant with $\lambda$ \cite{Brewczyk04,Brewczyk07}. 
We can identify $\lambda$ as a scaling of the Lieb-Liniger parameter $\gamma_{LL}$, since at any point in space
\be
\gamma_{LL}(\bo{x}) \propto g/|\phi(\bo{x})|^2 \propto \lambda^2. 
\ee
A single SGPE calculation represents a continuous family of systems with different $\gamma_{LL}$.

This symmetry is lost, as it must, in the PPR equations (\ref{PPR-eq}), whose dimensionless form is 
\bea\label{sc-ppsgpe}
i\frac{\partial\phi}{\partial t} = \\
(1-i\gamma)(H_{\rm sp}-\mu+g\phi\wt{\phi}^*)\phi + \sqrt{ig(1\!-\!2i\gamma)}\phi\xi + \sqrt{2\gamma T}\eta.\nonu
\eea
Here, while all the SGPE terms scale like $1/\sqrt{\lambda}\sim\gamma_{LL}^{-1/4}$, the magnitude of the quantum noise term is unchanged. This is how single-particle effects break the classical field description as $\gamma_{LL}$ grows from zero, and introduce a ``granularity'' that is inherently nonclassical. 

The $\gamma_{LL}$ parameter also has relevance to the accessible simulation time in PPR simulations of 1D systems, as follows: To encompass all the physics, such as the density fluctuations, one needs to have a numerical lattice that can resolve the inter-particle healing length (\ref{heall}). Hence, one needs $\Delta x\lesssim\xi_{\rm heal}\approx\hbar/\sqrt{mng}$. In a Thomas-Fermi approximation where $n(x)\approx[\mu-V(x)]/g$, the highest density $n_{TF}=\mu/g$ is the limiting case, so that we require $\Delta x\lesssim\hbar/\sqrt{m\mu}$. The timescale for physics occurring on the healing length-scale is 
\be\label{ut}
u_t=\frac{\hbar}{\mu},
\ee 
and from (\ref{sim_time}) one obtains that in 1D
\be
\frac{t_{\rm sim}}{u_t} \lesssim \frac{2}{(\gamma_{TF})^{1/6}}.
\label{sim_time_max}
\ee
 with 
\be
\gamma_{TF} = \frac{m g}{\hbar^2n_{TF}}
\ee
the  lowest value of $\gamma_{LL}(x)$, attained in the densest part of the cloud. 
This indicates that the equations (\ref{PPR-eq}) should be able to track processes related to the onset of inter-particle repulsion to their completion, provided we are in the regime when $\gamma_{TF}\ll1$. However, much slower processes such as thermalization in 1D will not reach saturation.

%%%%%%%%%%%%%%%%%%%%%%%%%%%%%%%%%%%%%%%%%%%%%%%%%%%%%%%%%%%%%%%%%
\section{Investigation of quantum granularity in a quench}
\label{RES}
%%%%%%%%%%%%%%%%%%%%%%%%%%%%%%%%%%%%%%%%%%%%%%%%%%%%%%%%%%%%%%%%%

We will investigate here the onset of quantum granularity and the effectiveness of the PPR equations (\ref{PPR-eq}) for describing it.
Since long time evolution and thermalization are ruled out for the reasons outlined above, to investigate the interplay between quantum and thermal fluctuations  we will take the following approach:
\begin{enumerate}
\item Evolve the SGPE (\ref{SGP_eq}) the same way as in Sec.~\ref{SGPEFLUCT} for a time $60/\omega$ to obtain a stationary ensemble of thermal states. This corresponds to a whole family of gases parametrized by $\gamma_{TF}$.
\item Input these samples into the PPR equations (\ref{PPR-eq}) explicitly choosing various values of $\gamma_{TF}$.
\item Evolve as long as possible and compare the resulting correlations to those described previously in Section.~\ref{SGPEFLUCT} for the SGPE. 
\end{enumerate}
The second point above implements an interaction quench. 
The idea is to have a quench that does not directly affect the cloud's mean-field properties and makes only small changes to the interaction energy. This aims to obtain a relatively clean display of the many-body effects of the quench, rather than more mundane effects that can be attributed to mean field evolution. Interaction quenches have been investigated for ultracold atom systems both in experiment \cite{Cheneau12,Trotzky12,Hung13} and theory, many with direct relevance to dilute 1D gases \cite{Carusotto10,Caux06,Caux07,Mossel12,Barmettler12,Rancon13,DeuarA,DeNardis14}. 

\subsection{Quench protocol}

Performing a quench directly in the manner of  (\ref{scaling}), and as calculated in \cite{Swislocki14}, is difficult experimentally. This is  because it is not straightforward to sufficiently rapidly change the linear density $n$ and even harder to simultaneously keep the density profile $|\phi(x)|^2$ unchanged or rapidly change the temperature in a uniform way. Instead of that, we can take advantage of an approximate scaling that occurs in the Thomas-Fermi regime (i.e. when $T\lesssim T_{\phi}$). Here, the density profile within the Thomas-Fermi radius $R_{TF}=(1/\omega)\sqrt{2\mu/m}$ is given by $n(x) = n(0)\left[1-(x/R_{TF})^2\right]$, while the chemical potential itself is $\mu\approx gn(0)$. Hence, the scaling
\bea
g &\to& \kappa\, g \nonu\\
\omega &\to& \sqrt{\kappa}\, \omega,\label{scaling2}\\
\mu &\to& \kappa\mu,\nonu
\eea
by a factor $\kappa$, while keeping temperature $T$ and density $n(x)$ constant, does not affect the Thomas-Fermi density profile. 
It does, however, affect the quantum granularity since  $\gamma_{LL} \propto \kappa$. Some small disturbance of the density profile near the classical turning points at $|x|\approx R_{TF}$ is to be expected.

This is a quench that \emph{can} be implemented by e.g. simultaneously increasing all trap frequencies by a factor of $\sqrt{\kappa}$. An increase of the transverse trapping frequency $\omega_{\perp}$ by this amount leads to a multiplication of $g$ by $\kappa$, since the latter is proportional to $\omega_{\perp}^2$ in 1D. 
What it does to the terms in Eq. (\ref{sc-ppsgpe}) is to multiply the deterministic part  by $\kappa$, the quantum noise by $\sqrt{\kappa}$, and the thermal noise is unchanged. Thus, the relative magnitude of quantum versus thermal noise grows with $\kappa$.

Quantities which remain unchanged under the scaling include the Thomas-Fermi radius $R_{TF}$, the phase coherence temperature $T_{\phi}$ of (\ref{Tphi}), the  central density $n(0)$, the temperature $T$, and the ideal gas critical temperature $T_c$, as well as all associated temperature ratios. On the other hand, neither the healing length $\xi_{\rm heal}$ of (\ref{heall}), nor the dimensionless interaction strength $\gamma_{TF}$, nor the ratio $T/\mu$ are invariant. 

\begin{figure}
\begin{center}
\includegraphics[width=\columnwidth]{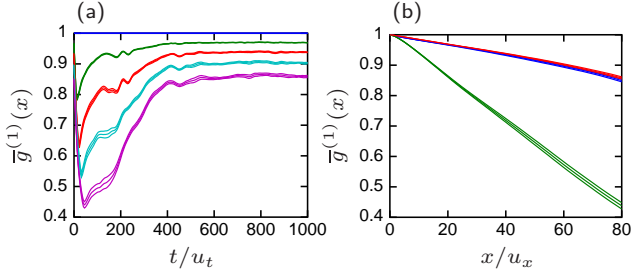}
\end{center}\vspace*{-4mm}
\caption[]{ 
Evolution of phase correlation $\wb{g}^{(1)}(x)$ in time under the SGPE after a quench (\ref{scaling2}) by a factor of $\kappa=20$. Time evolution in (a) ($x = 0, 20u_x, 40u_x, 60u_x, 80u_x$, descending) and spatial profile in (b) at times $0$ (blue), $45u_t$ (green), $1000u_t$ (red). Standard ``reference'' initial conditions with  $\mu=22.41$ and $T=0.156T_{\phi}$. 1$\sigma$ statistical uncertainty is shown as triple lines. Note, here the oscillator timescale is $1/\omega = 448u_t$.
\label{fig:SGP_quench}}
\end{figure}

The phase coherence length $L_{\phi}$ of (\ref{Thqcg1}) \emph{in equilibrium} is also preserved. However, we will see that this is not relevant for our quench, as the timescales for a reaction to the quench and rethermalization are very different. This can be seen from an SGPE calculation of the quench shown in figure~\ref{fig:SGP_quench}. Initially, $\wb{g}^{(1)}(r)$ undergoes a large change due to the quench, only to return to its initial values after a time of about $1/\omega$. 

The timescales accessible with the positive-P calculation do not reach the equilibration time, though. 
For this reason, the quantum fluctuation signal is not as clean as the (\ref{scaling}) quench described in \cite{Swislocki14}. It will be necessary to look at the \emph{difference} between c-field (SGPE ) calculations and the full quantum treatment of the positive-P simulation to study the effect of quantum fluctuations.

To generate initial conditions for the PPR, we will use the standard choices for atom fields  \cite{Deuar07}. 
If an initial state contains many atoms but is known only from its one-body wavefunction $\Psi_1(\bo{x})$, a close approximation is the coherent state with amplitude $\Psi_1(\bo{x})$. Then,  from the definition of the representation (\ref{ppmm}), one can immediately take
\be\label{psi1ic}
\Psi(\bo{x})=\wt{\Psi}(\bo{x})=\Psi_1(\bo{x}).
\ee
When the input state is described by a thermal ensemble (such as one generated by an SGPE, $\{ \Psi_{SGPE}(\bo{x}) \}$), an efficient choice is to generate one $j$th PPR sample for each $j$th SGPE sample $\Psi_{SGPE}^{(j)}(\bo{x})$, taking each such sample's one-body wavefunction as the input to the coherent initial condition (\ref{psi1ic}): 
\be\label{hybridic}
\Psi^{(j)}(\bo{x})=\wt{\Psi}^{(j)}(\bo{x})=\Psi^{(j)}_{SGPE}(\bo{x}).
\ee
This approach was used previously used e.g. in \cite{Kheruntsyan12} for initial conditions generated from a quasicondensate c-field ensemble via the expressions given in \cite{Petrov00,Dettmer01}.

%%%%%%%%%%%%%%%%%%%%%%%%%%%%%%%%%%%%%%%%%%%%%%%%%%%%%%%%%%%%%%%%%%%%%%%
\subsection{Emergence of quantum granularity with interaction strength}
%%%%%%%%%%%%%%%%%%%%%%%%%%%%%%%%%%%%%%%%%%%%%%%%%%%%%%%%%%%%%%%%%%%%%%%
\label{RES-g}

For the reference test case used in Sec.~\ref{SGPEFLUCT}, when $\mu=22.41$ and $g=0.1$, the interaction parameter is $\gamma_{TF}=0.00045$, indicating that we are still very deep in the semiclassical regime. 
We take the lowest temperature system of those described  in Sec.~\ref{SGPEFLUCT}, and vary the interaction strength and density in the positive-P simulation according to the scaling of (\ref{scaling2}).
Relative to the nominal case ($\mu=22.41, T=0.62\mu, g=0.1, \gamma=0.01, \wb{N}\approx2000$), we take values of $\kappa= 1, 5, 20$ which multiply the 1D interaction strength $g$ and change parameters as shown in table~\ref{table:g}. This increases the importance of quantum fluctuations as $\kappa$ rises. 
The simulation times achieved before excessive noise amplification set in, $t_{\rm sim}$, are also shown. They are of the same order as given by the expression (\ref{sim_time}). 

\newcommand{\om}{\hspace*{1em}}
\newcommand{\mom}{\hspace*{-1em}}
\begin{table*}
\caption{ 
Parameters for the simulated quenches with different levels of quantum fluctuations and temperatures. Here, as in the reference system of Sec.~\ref{SGPEFLUCT} which refers to an ${}^{87}{\rm Rb}$ gas, there are $N=2000$ atoms, and initial values of $g=0.1$ and $\mu=22.41$. $\nu_{\perp}=\omega_{\perp}/2\pi$, etc.
The $t_{\rm sim}$ are the maximum times reached. Plot color refers to Figs.~\ref{fig:densK_SGPvsPP_T60nK}-\ref{fig:g2_kappa20}, \ref{fig:g2cuts},and \ref{fig:nkmkxt}.
\label{table:g}}
\lineup

\begin{indented}
\item[]
\hspace*{2cm}\begin{tabular}{@{}ll@{\ }ll@{\ }cc@{\ }lll}
\br
$\kappa$	
		& $T$
				& 	$\,g$	&	$\mu$		&	$t_{\rm sim}/u_t$\mom	
													&	plot	&
														\multicolumn{3}{l}{reference system, $t>0$}\\
		&		&	\mom($t>0$)\om & ($t>0$)		&				&	\om\om color\om\om	&	 	$\nu$ [Hz]	&	$\nu_{\perp}$ [Hz]
																						& $T$ [nK] 	\\														
\mr
\01		&	13.9\om\om &	0.1	&	\022.41\om	&	5.7			&	magenta &	     \030.2		&	\0\0520	& 20	\\
\05		&	13.9	&	0.5	&	112.1\quad		&	3.8			&	cyan	&		156		&	\02600	& 20	\\
20		&	13.9	&	2.0	&	448.2		&	3.8			&	blue &		604		&	10400		& 20	\\
\mr
20		&	27.8	&	2.0	&	448.2		&	2.9			&	green &		604		&	10400		& 40	\\
20		&	42.8	&	2.0	&	448.2		&	2.9			&	red 	&		604		&	10400		& 62	\\
\br
\end{tabular}
\end{indented}
\end{table*}

\begin{figure}
\begin{center}
\includegraphics[width=0.75\columnwidth,clip]{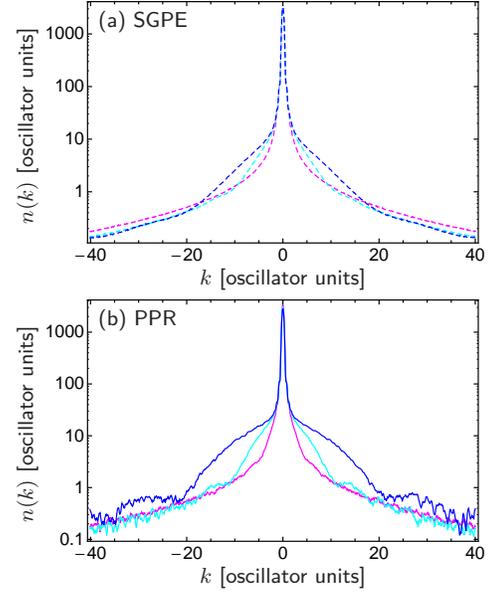}
\end{center}\vspace*{-4mm}
\caption[]{ Density in momentum space after $t=3u_t$, for different quench strengths  $\kappa=1$(no quench), $5, 20$ (colors as per table.~\ref{table:g}) starting from the $T=0.156T_{\phi}$ state. Panel (a): SGPE calculation, Panel (b): PPR calculation with quantum fluctuations. 
\label{fig:densK_SGPvsPP_T60nK}}
\end{figure}

Figure \ref{fig:densK_SGPvsPP_T60nK} shows the density in momentum space. The notable feature here is the appearance of additional scattered atoms in the wings of the distribution out to about $|k|\approx1/\xi_{\rm heal}$, the expected momentum corresponding to healing-length physics. The scattered number increases with $g$ as expected. Despite some quench physics occurring already in the SGPE, there are several times more scattered atoms in the full quantum PPR calculation due to quantum fluctuations, something whose effect will also be seen in other observables.

\begin{figure}
\begin{center}
\includegraphics[width=0.75\columnwidth,clip]{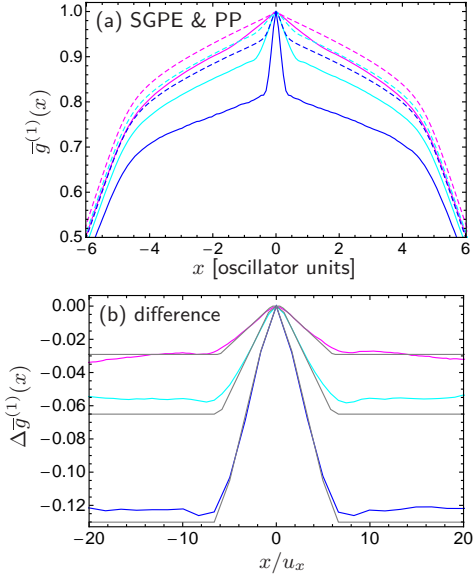}
\end{center}\vspace*{-4mm}
\caption[]{ 
Phase correlations $\wb{g}^{(1)}(x)$ at $t=3\,u_t$ after the quench from the $T=0.156T_{\phi}$ state. 
Panel (a) shows values calculated using the PPR (solid lines) and SGPE (dashed lines). Quench strengths were $\kappa=1$(no quench), $5, 20$ (color as per table.~\ref{table:g}). 
Panel (b) shows the difference in correlations $\Delta \wb{g}^{(1)}(x)$ due to the inclusion of quantum fluctuations. 
Gray lines show estimates (\ref{g1estquench}) based on the SGPE and a T = 0 quantum quench.
\label{fig:g1_T60nK}}
\end{figure}

\begin{figure}
\begin{center}
\includegraphics[width=0.75\columnwidth,clip]{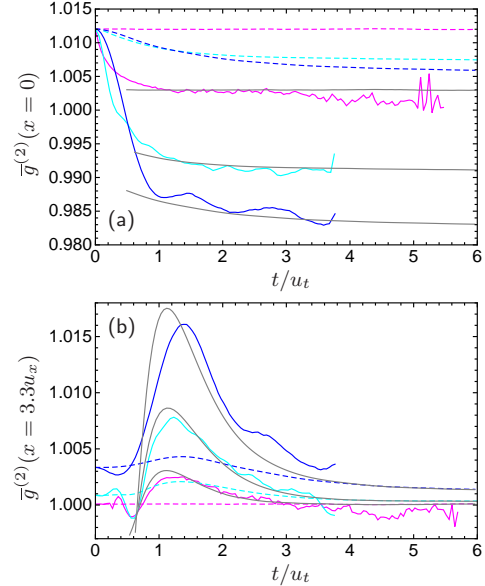}
\end{center}\vspace*{-4mm}
\caption[]{
Density correlations $\wb{g}^{(2)}(x)$ after the quench from the $T=0.156T_{\phi}$ state. Values calculated using the PPR (solid lines) and SGPE (dashed lines). Some high frequency statistical noise is seen in the PPR results at long times.
Panel (a) shows zero range correlations, while Panel (b) those at $x=3.3u_x$ with a correlation wave passing at times around $1.5u_t$. Quench strengths were $\kappa=1$(no quench), $5, 20$ (color as per table.~\ref{table:g}). 
Gray lines show estimates (\ref{g2estquench-0}) and (\ref{g2estquench-wave}) based on the SGPE and a $T = 0$ quantum quench.
\label{fig:g2_T60nK}}
\end{figure}

Figure \ref{fig:g1_T60nK} shows the averaged phase correlation function, $\wb{g}^{(1)}(x)$, in the center of the trap after an evolution time of $t=3u_t$, both for the SGPE and the full PPR treatment. The lower panel shows only the difference due to quantum fluctuations $\Delta\wb{g}^{(1)}(x)=\wb{g}^{(1)}_{\rm PP}(x)-\wb{g}^{(1)}_{\rm SGPE}(x)$. Figure \ref{fig:g2_T60nK} shows results for the corresponding density correlations, $\wb{g}^{(2)}(x)$, as a function of time. 
Despite the low values of the dimensionless interaction strength $\gamma_{TF}$ (having a maximum value of 0.0089 when $\kappa=20$), appreciable qualitative changes arise in the long-range properties of the gas due to quantum fluctuations. Phase coherence is reduced across all length scales, correlation waves are made stronger, and there is a reduction of the bunching. For sufficiently strong interactions, the desired antibunching appears on length scales of the order of $\xi_{\rm heal}$. All the effects grow in strength with $g$. 

In the figures~\ref{fig:densK_SGPvsPP_T60nK}--\ref{fig:g2_T60nK}, the difference between the magenta lines corresponding to $\kappa=1$ (``no quench'') shows the size of the transient introduced because equilibrium quantum fluctuations were not included in the initial SGPE-generated state. Its magnitude scales as $\sqrt{\gamma_{TF}}\sim\sqrt{g}$.

%%%%%%%%%%%%%%%%%%%%%%%%%%%%%%%%%%%%%%%%%%%%%%%%%%%%%%%%%%%%%%%%%%%%%%%
\subsection{Correlations as a function of temperature}
%%%%%%%%%%%%%%%%%%%%%%%%%%%%%%%%%%%%%%%%%%%%%%%%%%%%%%%%%%%%%%%%%%%%%%%
\label{RES-T}

The behavior of the difference due to quantum fluctuations bears close resemblance to recent predictions of correlation functions after a quantum quench of the interaction strength \cite{Carusotto10,Muth10,Barmettler12,DeuarA,Kormos14}.
We will now investigate it in more detail for a range of temperatures.  We choose the strongest $\kappa=20$ quench to heighten the visibility of quantum fluctuation effects. Temperatures correspond to the three SGPE calculations in Sec.~\ref{SGPEFLUCT}, describing for example ${}^{87}$Rb in the traps and temperatures given in table~\ref{table:g}. 

\begin{figure}
\begin{center}
\includegraphics[width=0.75\columnwidth,clip]{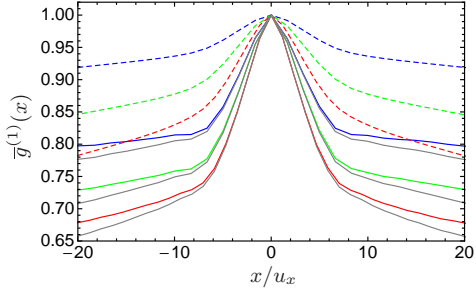}
\end{center}\vspace*{-4mm}
\caption[]{ 
Phase correlations $\wb{g}^{(1)}(x)$ at $t=3\,u_t$ after the $\kappa=20$ quench, at three values of temperature matching the SGPE results of figure~\ref{fig:g1_SGP_u}: $T=0.156T_{\phi}$ (blue), $T=0.311T_{\phi}$ (green), and $T=0.480T_{\phi}$ (red). Values calculated using the PPR (solid lines) and SGPE (dashed lines).
The gray lines show estimates (\ref{g1estquench}) based on the SGPE and a $T=0$ quantum quench.
\label{fig:g1_gx20_TxnK_u}}
\end{figure}

\begin{figure}
\begin{center}
\includegraphics[width=0.75\columnwidth,clip]{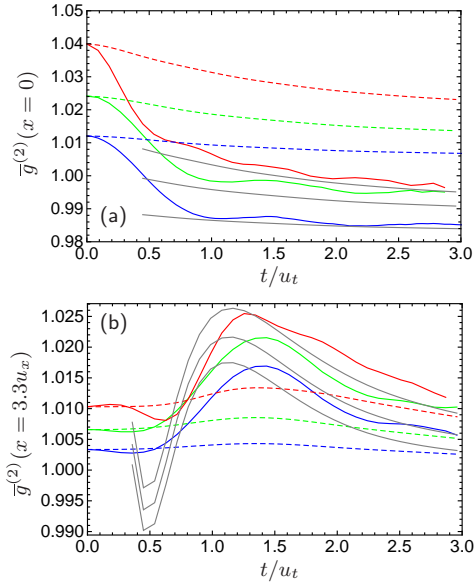}
\end{center}\vspace*{-4mm}
\caption[]{ 
Density correlations $\wb{g}^{(2)}(0)$ after a $\kappa=20$ quench, at three values of temperature matching the SGPE results of figure~\ref{fig:g2_SGP_u}: $T=0.156T_{\phi}$ (blue), $T=0.311T_{\phi}$ (green), and $T=0.480T_{\phi}$ (red).
Values calculated using the PPR (solid lines) and SGPE (dashed lines).
Panel (a) shows zero range correlations, while Panel (b) those at $x=3.3u_x$ with a correlation wave passing at times of around $1.5u_t$.  The gray lines show estimates (\ref{g2estquench-0}) and (\ref{g2estquench-wave}) based on the SGPE and a $T = 0$ quantum quench.
\label{fig:g2_kappa20}}
\end{figure}

The correlations are shown in Figs.~\ref{fig:g1_gx20_TxnK_u} and~\ref{fig:g2_kappa20}. 
Qualitatively, the quantum fluctuations are seen to add to the existing thermal behavior in the SGPE. That is,  there is additional phase decoherence, while for density fluctuations there is a transition between bunched behavior and antibunching when the temperature is low enough, as expected from the full quantum physics. 

Quantitatively, the quench-like behavior turns out to be well approximated by adding the $T=0$ predictions for dilute gases found in \cite{DeuarA} and thermal effects seen in the plain SGPE. The rough estimates for medium and long times $t\gtrsim u_t$ are shown in figures~\ref{fig:g1_T60nK}--\ref{fig:g2_kappa20} as grey lines.  For phase fluctuations, they are:
\bea\label{g1estquench}
g^{(1)}_{\rm est}(x) = \lefteqn{g^{(1)}_{\rm SGPE}(x)}\!\!\!\!\!\!\\ -\frac{\sqrt{\gamma_{TF}}}{8}\times\left\{\begin{array}{c@{\quad\text{if}\quad}l}0 & x < \xi_{\rm heal}/2 \\2x/\xi_{\rm heal}-1 & \xi_{\rm heal}/2<x<2t\xi_{\rm heal}/u_t\\4t/u_t-1&\xi_{\rm heal}>2t\xi_{\rm heal}/u_t\end{array}\right.
\nonumber\eea
The density fluctuation estimate is
\numparts
%\label{g2estquench}
\be\label{g2estquench-0}
g^{(2)}_{\rm est}(x) = g^{(2)}_{\rm SGPE}(x,t) -\frac{\sqrt{\gamma_{TF}}}{2} C_k e^{-2x/\xi_{\rm heal}} 
\ee
for small $x\sim\mc{O}(\xi_{\rm heal})$, and 
\bea\label{g2estquench-wave}
&&g^{(2)}_{\rm est}(x,t) = \\&&g^{(2)}_{\rm SGPE}(x,t) +\frac{\sqrt{\gamma_{TF}}}{2}\left(\frac{u_t}{6t}\right)^{\frac{1}{3}}{\rm Ai}\left[\left(\frac{4u_t}{3t}\right)^{\frac{1}{3}}\left(\frac{2t}{u_t} - \frac{x}{\xi_{\rm heal}}\right)\right]\nonumber
\eea
for large distances $x\gg\xi_{\rm heal}$. Here, ${\rm Ai}[x]$ is the Airy function, and $C_k$ is a constant that is unity in a continuum system and 
\be\label{Kdef}
C_k = \frac{2}{\pi} \,\tan^{-1}\left[\frac{k_{\rm max}\xi_{\rm heal}}{2}\right]
\ee  
\endnumparts
when a lattice wavevector cutoff $k_{\rm max}$ is present.
The first estimate gives the antibunching dip (or the reduction of bunching at higher temperatures), while the second gives the additional correlation wave intensity.

%%%%%%%%%%%%%%%%%%%%%%%%%%%%%%%%%%%%%%%%%%%%%%%%%%%%%%%%%%%%%%%%%%%%%%%
\subsection{Pairs in momentum space}
%%%%%%%%%%%%%%%%%%%%%%%%%%%%%%%%%%%%%%%%%%%%%%%%%%%%%%%%%%%%%%%%%%%%%%%
\label{RES-k}

\begin{figure}
\begin{center}
\includegraphics[width=\columnwidth,clip]{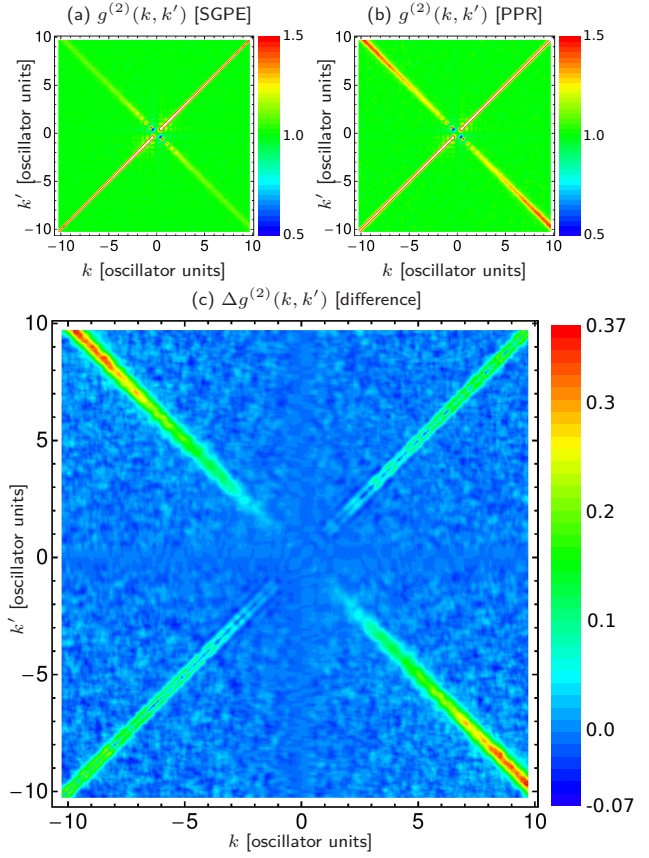}
\end{center}\vspace*{-4mm}
\caption[]{
Correlation function $g^{(2)}(k,k')$ at $t=1.0u_t$ after a $\kappa=20$ quench with the  $T=0.156T_{\phi}$ initial condition.
Panel (a): shows the results of an SGPE calculation, Panel (b) of the full PPR evolution, and Panel (c) the difference.  
White color in top panels indicates $g^{(2)}(k,k')>1.5$. 
\label{fig:g2kki}}
\end{figure}

As mentioned at the end of Sec.~\ref{SGPEFLUCT}, one expects to see pairing in momentum space due to quantum fluctuations of Bogoliubov phonons. 
The baseline SGPE behavior of $g^{(2)}(k,k')$ is shown in figure~\ref{fig:g2kki}a. We use  the lowest temperature $T=0.156T_{\phi}$. It shows a HBT thermal fluctuation peak along the $k\approx k'$ line and 
the condensate correlation behavior discussed in Sec.~\ref{SGPEFLUCT} at small momenta $|k|,|k'|\lesssim1$. Some pairing $k'\approx-k$ is also seen. 
The corresponding result of the full PPR simulation is shown in figure~\ref{fig:g2kki}b, and the difference between them in figure~\ref{fig:g2kki}c. 
The quantum fluctuations introduce significantly more pairing  between counter-propagating atoms ($k'\approx -k$), particularly at large momenta, greater than those spanned by the condensate. There is also a broadening of the HBT correlations due to quantum fluctuations seen as the double diagonal line in figure~\ref{fig:g2kki}c.

\begin{figure}
\begin{center}
\includegraphics[width=0.75\columnwidth,clip]{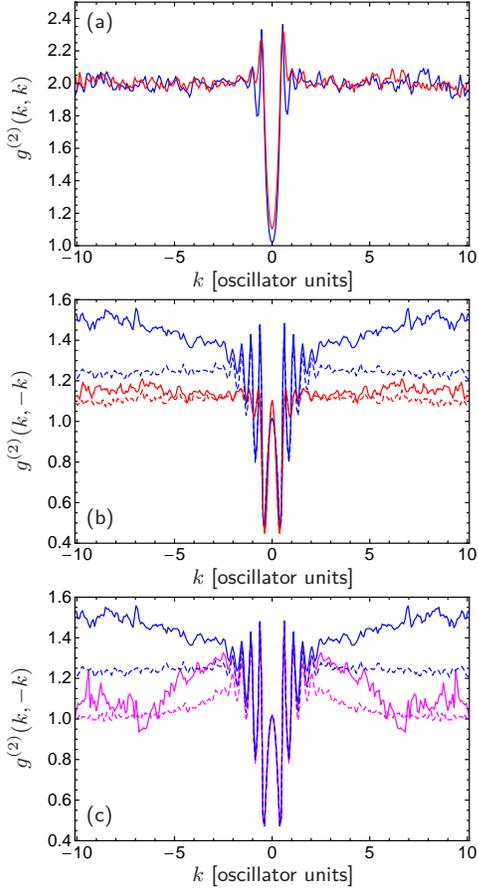}
\end{center}\vspace*{-4mm}
\caption[]{ 
Slices through the momentum correlation function after $t=2.0u_t$ evolution with the PPR. Panels (a) and (b) show the temperature variation of co-propagating $g^{(2)}(k,k)$
and counter-propagating pair correlation $g^{(2)}(k,-k)$, respectively. Colors as in table~\ref{table:g}. 
Panel (c) shows the dependence of the counter-propagating pair correlations  $g^{(2)}(k,-k)$ on $g$, for different quench strengths. 
Corresponding SGPE results shown as dashed lines. Noise at large $k$ values is statistical; 10 000 realizations were used.
\label{fig:g2cuts}}
\end{figure}

Further details are shown in figure~\ref{fig:g2cuts}. Panels (a) and~(b) show cuts along $k'=k$ and $k'=-k$, respectively, for two of the temperatures we have been considering. The pair correlation rises across a wide range of momenta as temperature drops, while the HBT fluctuation peak in Panel (a) is unaffected. 
Panel (c) of figure~\ref{fig:g2cuts} shows the increase of pairing with $\kappa$.

\begin{figure}
\begin{center}
\includegraphics[width=\columnwidth,clip]{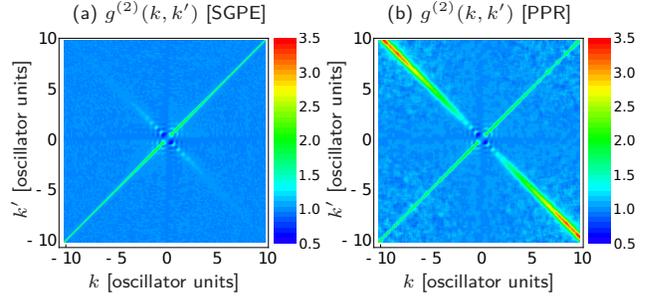}
\end{center}\vspace*{-4mm}
\caption[]{ 
Correlation function $g^{(2)}(k,k')$ after $t=0.3u_t$ of evolution subsequent to a $\lambda=20$ quench of the ``clean'' (\ref{scaling}) type, using the  $T=0.156T_{\phi}$ initial condition.
Panel (a): shows the results of an SGPE calculation, Panel (b) of the full PPR evolution.  Note the absence of high momentum pairs in the SGPE quench.
\label{fig:cleanquench}}
\end{figure}

In the clean, but not very physical quench (\ref{scaling}), counter-propagating pairs are \emph{only} produced by quantum fluctuations as shown in figure~\ref{fig:cleanquench}. For the physical quench (\ref{scaling2}), however, an additional classical correlation between counter-propagating waves is already induced by the quench without requiring discrete pair production.

Inspection of Figs.~\ref{fig:g2kki}c and \ref{fig:g2cuts} allows us to assess physically whether the pairs in the trapped gas can act as a source of nonclassical atom pairs when they are released from the trap. 
For example, in experiments with BEC collisions, released atoms were binned in momentum, and the distributions of bin occupations analyzed to show sub-Poissonian number fluctuations (number squeezing) and Cauchy-Schwartz inequality violation \cite{Jaskula10,Kheruntsyan12}. It was found that for either effect to be present, one needs bin averaged $g^{(2)}(k,k')$ with $k'$ and $k$ in different bins to be larger than the $g^{(2)}(k,k')$ averaged in a single bin. In our case here, one would take $k$ intervals on either side of the condensate as bins. Looking at the figures, the pair ($k'\approx-k$) and local density ($k'\approx k$) correlations have heights of about 1.5 and 2, respectively, and similar peak widths. We conclude that it is not possible to obtain released nonclassical atom pairs for our parameters because the in-situ pairs are not sufficiently correlated.

%%%%%%%%%%%%%%%%%%%%%%%%%%%%%%%%%%%%%%%%%%%%%%%%%%%%%%%%%%%%%%%%%%%%%%%
\subsection{Resulting stationary state}
%%%%%%%%%%%%%%%%%%%%%%%%%%%%%%%%%%%%%%%%%%%%%%%%%%%%%%%%%%%%%%%%%%%%%%%
\label{RES-t}

Despite the simulation time limitations (\ref{sim_time}) in the PPR equations, some observable quantities reach stable values, at least on the timescales studied.

\begin{figure}
\begin{center}
\includegraphics[width=\columnwidth]{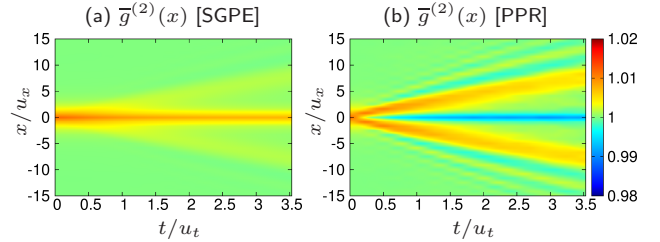}
\end{center}\vspace*{-4mm}
\caption{ 
Time dependence of the density correlation function $\wb{g}^{(2)}(x)$ after a $\kappa=5$ quench from the  $T=0.156T_{\phi}$ initial condition.
Panel (a): SGPE, Panel (b): full PPR evolution. Antibunching appears and stabilizes near $x=0$. 
\label{fig:g2xt} }
\end{figure}

\begin{figure}
\begin{center}
\includegraphics[width=\columnwidth]{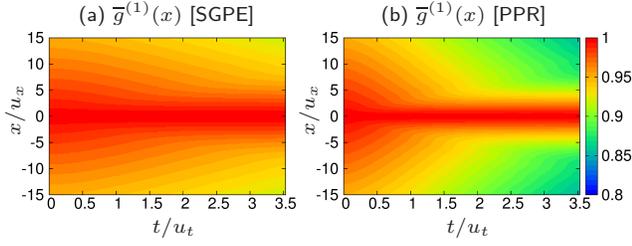}
\end{center}\vspace*{-4mm}
\caption{
Time dependence of the phase correlation function $\wb{g}^{(1)}(x)$ after a $\kappa=5$ quench from the  $T=0.156T_{\phi}$ initial condition.
Panel (a): SGPE, Panel (b): full PPR evolution. 
\label{fig:g1xt} }
\end{figure}

\begin{figure}
\begin{center}
\includegraphics[width=0.75\columnwidth]{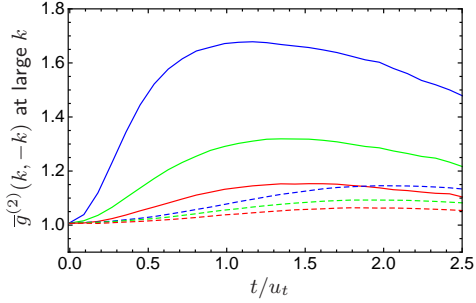}
\end{center}\vspace*{-4mm}
\caption{ 
Time evolution of the pairing correlation at large $k$ after a $\kappa=20$ quench for $T=0.156T_{\phi}$ (blue), $T=0.311T_{\phi}$ (green), and $T=0.480T_{\phi}$ (red) initial states.
The plot shows the peak value $g^{(2)}(k,-k)$ after averaging over the range $k\in[0.5,1]/\xi_{\rm heal} = [10.6,21.2]$ to give $\wb{g}^{(2)}(k,-k)$ and improve the signal-to-noise ratio. 
Solid lines: full PPR evolution, dashed: SGPE.
\label{fig:nkmkxt}
}
\end{figure}

The full quantum evolution of density correlations is shown in Figs.~\ref{fig:g2_T60nK}, \ref{fig:g2_kappa20} and~\ref{fig:g2xt}(b). Note the settling of the local bunching/antibunching to a stationary value in figure~\ref{fig:g2_T60nK}. For more long-range correlations, one observes quite long-lived waves moving away from the small $x$ region, whereas locally only the stationary antibunching remains. 
Stabilization of short-range correlations over a progressively larger region with time is also seen in the phase correlations, which are shown in figure~\ref{fig:g1xt}. There, one can see the initial reduction of phase coherence due to quantum fluctuations, and later a changeover to a stable profile that is seen as a kink in the color contours. 
The appearance of counter-propagating pairs is shown in figure~\ref{fig:nkmkxt} for large momentum, the region in which pairs dominate other effects.

The late-time stationary state has the qualitative features expected of a fully quantum thermal equilibrium state: antibunching, increased phase decoherence, an increase in counter-propagating pairs like in a Bogoliubov description. On the other hand, obtaining the thermal equilibrium would, in fact, be surprising since the timescale of a few $u_t$ is too short to thermalize energy differences much smaller than $\mu$, e.g. those involved in long-wavelength phase-fluctuations. This is reflected in the ongoing evolution of $g^{(1)}(x)$  at large $x$, seen in figure~\ref{fig:g1xt}.

The density self-correlation $g^{(2)}(0)$ after the clean ``$\lambda$'' quench is well suited for a precise investigation of this from a theoretical angle, provided the quantum depletion in the initial state is very small. To satisfy the latter condition, we use a set of SGPE initial conditions rescaled by (\ref{scaling}) with respect to the ``reference'' case so that the initial interaction strength is $g=0.01$. The size of the remaining transient in $g^{(2)}(0)$ is the difference between the last and 5th column in table~\ref{table:g2}, in this case $\approx0.001$. It would be $\approx0.009$ without the rescaling, as seen in the $\kappa=1$ (solid magenta) line of figure~\ref{fig:g2_T60nK}(a).
Values obtained with the SGPE and full PPR equations are compared in table~\ref{table:g2} to the exact quantum thermal equilibrium value obtained for the uniform gas by Yang \& Yang \cite{Yang69}, and some estimates. 
Estimates are simpler here because unlike the ``$\kappa$'' quench, the thermal baseline remains the same as at $t=0$.  
The first two columns regarding $g^{(2)}(0)$ show that the SGPE is well matched by the thermal fluctuation estimate (\ref{Thqcg2}). 
The last two show very good agreement between the stationary state and the quench + thermal fluctuations estimate (\ref{g2estquench-0}). However, the degree of antibunching in the exact quantum equilibrium result is appreciably greater than in the quench and PPR simulations. Indeed, in the limit of small values of $\gamma_{TF}$, 
the quench reduces $g^{(2)}(0)$ by $(C_k/2)\sqrt{\gamma_{TF}}$, which is $\le\half\sqrt{\gamma_{TF}}$, while the reduction in the exact quantum equilibrium state is  $(2/\pi)\sqrt{\gamma_{TF}}$ \cite{Kheruntsyan03}, i.e at least 27\% larger. 

The stationarity of the evolution within the sound cone in Figs.~\ref{fig:g2_T60nK}, \ref{fig:g2xt}, and \ref{fig:g1xt} shows that
any later equilibration there is negligible on $u_t$ timescales despite the scattered particles interacting with each other and the remainder of the system. This can be considered another case of ``pre-thermalization'' \cite{Berges04,Rigol09,Gring12,DeNardis14,Kollath07}.

\begin{table}
\caption{ 
Comparison of stationary density self-correlation $g^{(2)}(0)$ values in simulations of the clean ``$\lambda$'' quench with relevant estimates (``est.'') and thermal ensemble values. 
The calculated $\wb{g}^{(2)}(0)$ values (``calc.'') come from SGPE (\ref{SGP_eq}) and PPR (\ref{PPR-eq}) simulations at the final times given by $t/u_t$ in the table.
The true thermal equilibrium  values are from a calculation of the exact Yang \& Yang solution \cite{Yang69} using $\gamma_{TF}$ based on the central density in the Thomas-Fermi approximation, $n_{TF}=\mu/g$.
In all cases, $\mu=22.41$ and  $g=0.01\lambda$ to reduce spurious transients, as explained in the text.
The upper part shows variation with $g$, keeping the SGPE relative temperature $T/T_{\phi}=0.156$ constant, while the lower part shows variation with $T/T_{\phi}$, keeping interaction strength $g=0.2$ and $\gamma_{TF}=0.001785$ constant.
The statistical uncertainty for the numerical calculations is $\sim\pm0.0001$ for the SGPE, and up to $\sim\pm0.001$ for PPR at large $\lambda$.
The numerical lattice used had $k_{\rm max} = 8.472/\xi_{\rm heal}$ for all cases, except for the last line, where $k_{\rm max} = 11.981/\xi_{\rm heal}$.
\label{table:g2}}
\lineup
\begin{indented}
\item[]
\begin{tabular}{@{}llll@{\ }ll@{\ }l@{\ }ll}
\br
	&		&		&		&	\multicolumn{5}{c}{ density self-correlations $g^{(2)}(0)$ } 									\\
\cline{5-9}
	&		&  		& 		&	\multicolumn{2}{c}{thermal only}		&	\multicolumn{3}{l}{ + quantum fluctuations}				\\
\cline{5-9}
$g$	& $\lambda$	& $T/\mu$	& $t/u_t$\om &	calc.		&	thermal		& 	exact			&	calc.			& 	quench 			\\
	&		&		& 		&	via		&	est. by		&	result		&	via 			&	est. by			\\
	&		& 		&   		&	SGPE		&	(\ref{Thqcg2})	&	\cite{Yang69}	&	PPR			&	(\ref{g2estquench-0})	\\
\mr
0.01\mom\mom&\01	& 0.156	& 20.5	&	1.0120	&	1.0131		&	1.0084		&	1.0108		&	1.0111			\\
0.05\mom\mom&\05	& 0.156	& \07.2	&	1.0120	&	1.0131		&	0.9982		&	1.0068		&	1.0075			\\
0.1	&	10	& 0.156	& \05.6	&	1.0120	&	1.0131		&	0.9899		&	1.0015		&	1.0030			\\
0.2	&	20	& 0.156	& \03.5	&	1.0120	&	1.0131		&	0.9753		&	0.9923		&	0.9940			\\
0.4	&	40	& 0.156	& \02.0	&	1.0120	&	1.0131		&	0.9482		&	0.9702		&	0.9760			\\
\mr
0.2	&	20	& 0.156	& \02.7	&	1.0120	&	1.0131		&	0.9753		&	0.9923		&	0.9940			\\
0.2	&	20	& 0.312	& \02.7	&	1.0240	&	1.0262		&	0.9796		&	1.0045		&	1.0060			\\
0.2	&	20	& 0.480	& \02.7	&	1.0399	&	1.0404		&	0.9858		&	1.0189		&	1.0210			\\
\br
\end{tabular}
\end{indented}
\end{table}

%%%%%%%%%%%%%%%%%%%%%%%%%%%%%%%%%%%%%%%%%%%%%%%%%%%%%%%%%%%%%%%%%%%%%
\section{Conclusions}
%%%%%%%%%%%%%%%%%%%%%%%%%%%%%%%%%%%%%%%%%%%%%%%%%%%%%%%%%%%%%%%%%%%%%

We have derived the positive-P equations for the PSGPE (\ref{ppr-eq}) and SGPE (\ref{PPR-eq}) models. Treating c-field states at $T>0$ this way does indeed generate the expected types of quantum fluctuation phenomena, and integrates them on an equal footing with thermal fluctuations. One sees the appearance of antibunching (or a reduction of bunching), additional reduction of phase coherence in comparison with purely thermal phase fluctuations, and correlated atom pairs with opposite momenta in situ in the trapped cloud. Quantum fluctuations effects can be large, even at ``warm'' temperatures that are too high for a Bogoliubov description, e.g. $T\approx 0.4T_{\phi}$. 

In practice, the leading inaccuracy in our test calculations came from the lack of built-in quantum depletion in the c-field initial conditions. Depletion is subsequently built by a transient process at early times $t>0$ by the equations. This can be either an important or only a minor issue, depending on the problem. For example, this contribution can be seen in figure~\ref{fig:g1_T60nK} as the magenta line ($\kappa=1$) that eventuates when there is no change in the Hamiltonian but only in the equations. For strong quenches, $\kappa\gg1$, the transient contribution due to the initial state becomes small in comparison with the new quantum fluctuations produced as a result of the quench. For low enough initial temperatures, better initial state quantum fluctuations could be generated in the Bogoliubov treatment, and then evolved using (\ref{PPR-eq}) even into regimes where the Bogoliubov approximation ceases to apply. However, the generation of truly equilibrium quantum fluctuations in a gas with small condensate fraction is difficult, and remains a ``holy grail'' of sorts. 

Remembering to keep an eye on the initial quantum depletion issue, the Equations (\ref{PPR-eq}) could be used to treat physical phenomena that occur in non-condensates on timescales compatible with the estimate (\ref{sim_time}). 
The available time is often sufficient to stabilize observables to their metastable values -- in particular, correlations within the ``sound cone'', and especially the antibunching $g^{(2)}(0)$ and quantum depletion contribution to phase correlations $g^{(1)}(x)$. In contrast to previous work using a stochastic Bogoliubov approach \cite{Krachmalnicoff10,Deuar11}, atoms scattered to modes that are \emph{not} strongly separated from the source cloud are not a problem here. The equations could also be used to generate initial conditions with quantum depletion by evolving to the quasi-stationary state, although the amount of depletion is not exactly the same as in thermal equilibrium. Note that the long equilibration time for some observables may mean that physical clouds are not always in thermal equilibrium, anyhow. 

The approach is applicable to nonuniform, inhomogeneous gases, and time-dependent Hamiltonians, because it relies on stochastic equations in a simple position basis space. Like other positive-P representation based methods, the computational complexity scales linearly with the number of modes, allowing equally well for 1D as well as 2D and 3D systems. The equations that we use have a different structure than other recent approaches treating spontaneous processes at nonzero temperature because they do not introduce a separation between source and scattered modes like in stochastic Bogoliubov expansions of c-fields (see \cite{Kheruntsyan12,Wasak12}) or a separation between differently treated Wigner and PPR modes as in \cite{Hoffmann08}. They allow for interaction between all modes to all orders in the same manner, but with a simulation time price.  

The example calculations with large $\kappa$ have realized basically a quantum quench at nonzero temperature, and demonstrated how thermal and quantum fluctuations phenomena come to coexist. 
They indicate that in many cases the $T>0$ behavior can be modeled by a simple addition of $T=0$ quench results and thermal c-field calculations. Very high temperatures with $n_0\to0$ were not yest investigated, however. In particular, we see that the degree of antibunching in the metastable state given by $g^{(2)}(0)$ is significantly weaker (by about a third) than the equilibrium value. This has consequences for the later dynamics and energy balance of the gas, because the interaction energy is directly proportional to $g^{(2)}(0)$. It does not relax to its equilibrium value on the seemingly obvious timescale of $1/gn$ that corresponds to the interaction energy per particle, but much slower.

%%%%%%%%%%%%%%%%%%%%%%%%%%%%%%%%%%%%%%%%%%%%%%%%%%%%%%%%%%%%%%%%%%%%%
\ack
%%%%%%%%%%%%%%%%%%%%%%%%%%%%%%%%%%%%%%%%%%%%%%%%%%%%%%%%%%%%%%%%%%%%%
We are grateful to Nikolaos Proukakis, Stuart Cockburn, and Donatello Galucci for helpful discussions. This work was
 supported by the National Science Centre (Poland) Grant No. 2012/07/E/ST2/01389. We also acknowledge support by the Marie Curie European Reintegration Grant PERG06-GA-2009-256291 and by the Polish Government project 1697/7PRUE/2010/7 during initial exploratory work.

%%%%%%%%%%%%%%%%%%%%%%%%%%%%%%%%%%%%%%%%%%%%%%%%%%%%%%%%%%%%%%%%%%%%%
\section*{References}
\bibliography{SGP}
%%%%%%%%%%%%%%%%%%%%%%%%%%%%%%%%%%%%%%%%%%%%%%%%%%%%%%%%%%%%%%%%%%%%%

\end{document}